\documentclass[aps,prxquantum,reprint,superscriptaddress,footinbib]{revtex4-2}

\DeclareUnicodeCharacter{2212}{-}
\usepackage{braket}
\usepackage[utf8]{inputenc}
\usepackage{csquotes}
\usepackage{xcolor}
\usepackage{graphicx} 
\usepackage{epsfig}
\usepackage{epstopdf}
\usepackage{braket}
\usepackage[normalem]{ulem}
\usepackage[colorlinks]{hyperref}
\usepackage{bm}
\usepackage{siunitx}
\DeclareSIUnit\gauss{G}
\DeclareSIUnit\rad{rad}	
\usepackage{amsmath}
\usepackage{bbold}
\usepackage{orcidlink}

\usepackage{soul}

\def\-{\raisebox{.75pt}{-}}

\usepackage{ulem}

\newcommand{\Ca}{^{40}\text{Ca}^+}

\newcommand{\Ss}{4^2\mathrm{S}_{1/2}}
\newcommand{\Pp}{4^2\mathrm{P}_{3/2}}
\newcommand{\Ppo}{4^2\mathrm{P}_{1/2}}
\newcommand{\Dd}{3^2\mathrm{D}_{5/2}}
\newcommand{\Ddt}{3^2\mathrm{D}_{3/2}}

\usepackage{placeins}

\begin{document}

\title{A Cavity-Interfaced Register of Trapped-Ion Qubits with Multi-Second Coherence}%

\author{J.~Bate\,\orcidlink{0000-0002-3570-5102}}
\affiliation{Universit\"at Innsbruck, Institut f\"ur Experimentalphysik, Technikerstr. 25, 6020 Innsbruck, Austria}
\author{J.~Helgert\,}
\affiliation{Universit\"at Innsbruck, Institut f\"ur Experimentalphysik, Technikerstr. 25, 6020 Innsbruck, Austria}
\affiliation{LaserLaB, Department of Physics and Astronomy, Vrije Universiteit Amsterdam, De Boelelaan 1100, 1081 HZ Amsterdam, Netherlands}
\author{M.~Canteri\,\orcidlink{0000-0001-9726-2434}}
\affiliation{Universit\"at Innsbruck, Institut f\"ur Experimentalphysik, Technikerstr. 25, 6020 Innsbruck, Austria}
\author{V.~Krutyanskiy\,\orcidlink{0000-0003-0620-4648}}
\affiliation{Universit\"at Innsbruck, Institut f\"ur Experimentalphysik, Technikerstr. 25, 6020 Innsbruck, Austria}
\author{B.~P.~Lanyon\,\orcidlink{0000-0002-7379-4572}}
\affiliation{Universit\"at Innsbruck, Institut f\"ur Experimentalphysik, Technikerstr. 25, 6020 Innsbruck, Austria}
\email[Correspondence should be sent to ]{ben.lanyon@uibk.ac.at}

\date{\today}

\begin{abstract}

Scalable quantum networks require generating remote entanglement faster than decoherence erases it, calling for nodes that combine efficient light–matter interfaces with long coherence times. 
Cavity-coupled trapped ions are a promising platform for network nodes, offering high-efficiency extraction of photons. Here, we report multi-second coherence for a qubit register in a cavity-integrated ion trap.
First, using dynamical decoupling on spin ground states, the coherence times of five co-trapped ion qubits are extended into the multi-second regime. 
Second, cavity-collected ion-photon entanglement is faithfully stored in these protected ion-memory states for multiple seconds. 
Third, we show that ion‑memory qubits can be robust to the generation of over a thousand cavity photons by a co‑trapped ion, with decoherence limited by laser crosstalk.
Finally, we estimate that a modestly improved and duplicated ion-cavity system could enable the simultaneous establishment of multiple remote Bell pairs between two remote ion registers, representing a step toward networking multi-qubit prototype trapped-ion quantum processors. 
\end{abstract}

\maketitle

Envisioned quantum networks consist of distributed nodes connected by photonic links \cite{Kimble2008, doi:10.1126/science.aam9288, Covey:2023qy}.
The nodes contain registers of matter-based qubits for the storage and processing of quantum information, while the photonic links distribute entanglement between nodes via travelling photons.
When deployed on the single-laboratory scale, such networks could offer a scalable modular approach to quantum computing and simulation \cite{PhysRevA.89.022317}. Over larger distances, such networks offer powerful new applications for enhanced sensing~\cite{dur, 3hgx-wcdn, Giovannetti2011}, timekeeping~\cite{networkclocks, Marciniak:2021tld}, communication~\cite{PhysRevLett.67.661}, security~\cite{Bugalho2025privaterobuststates,security}, and cloud-based computing~\cite{Fitzsimons:2016xat}.  
State-of-the-art experiments have demonstrated prototype quantum networks involving the remote entanglement of two \cite{Saha2025,PhysRevLett.130.090803,PhysRevLett.125.260502,PhysRevLett.119.010503, Ritter2012} and three \cite{doi:10.1126/science.abg1919, seubert2026efficiententanglementremotesingleatom} matter-based qubits in a laboratory setting, and in some cases between distant buildings \cite{NetworkPaper, Knaut2024,  Cui:2025olz, Stolk:2024xop, vanLeent2022, Hensen2015}. %
A current challenge is to develop network node architectures that combine multi-qubit registers, universal quantum processing, efficient photon interfaces and long qubit coherence times.

\begin{figure}[t!]
	\begin{center}
       \includegraphics[width=0.95\columnwidth]{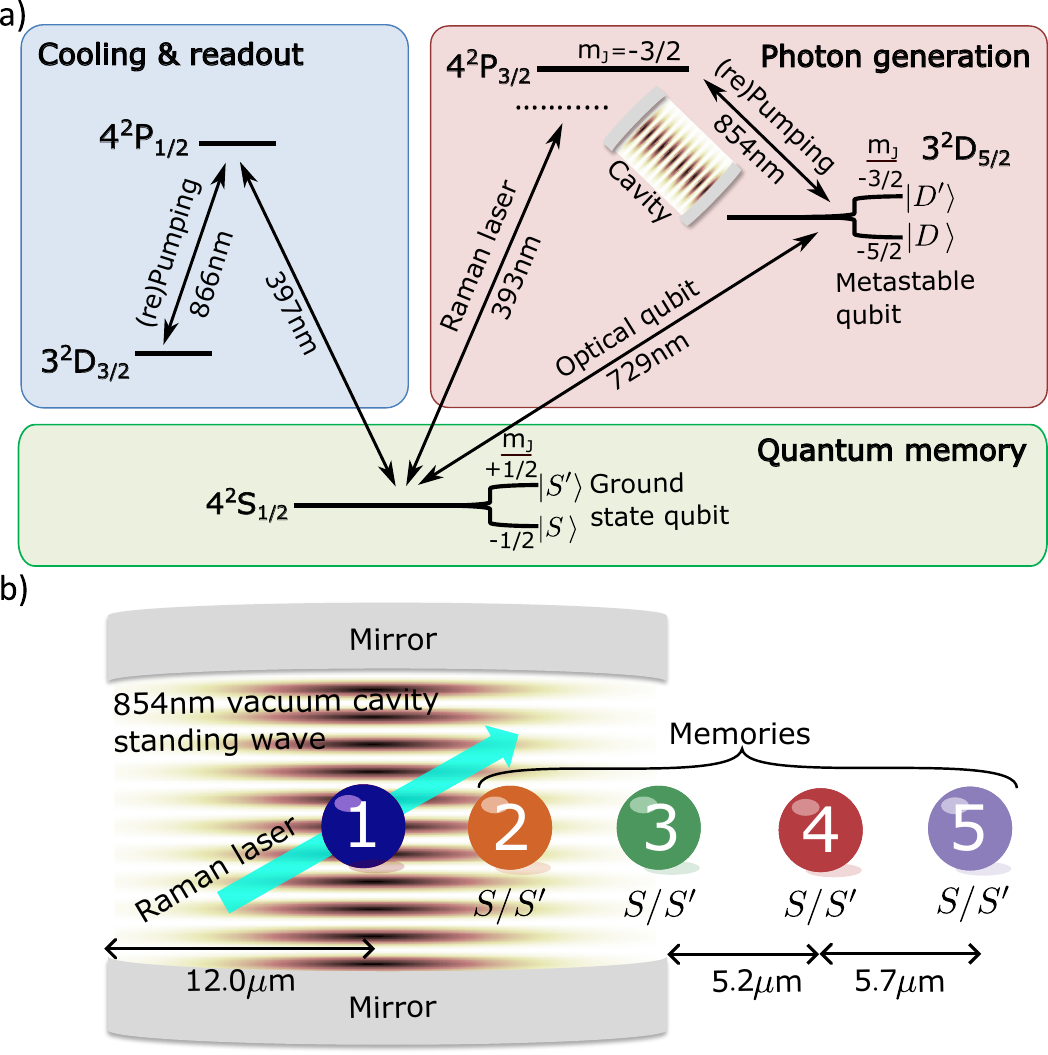}
		\caption{
\label{fig:intro} Cavity-interfaced multi-qubit memory register.
(a) Energy-level diagram of $^{40}$Ca$^+$ showing the relevant transitions for photon generation and qubit encodings.
(b) Schematic of five co-trapped $^{40}$Ca$^+$ ions (trap electrodes not shown). A cavity photon is generated from ion 1 via a focused \SI{393}{\nano\meter} Raman pulse. The remaining ions serve as quantum memories, with qubits encoded in the ground-state spin levels ($\ket{S}$/$\ket{S'}$). The cavity mode has a waist of \SI{12.0}{\micro\meter}. The shown ion separations are symmetric around ion 3. }
		\vspace{-3 mm}
	\end{center}
\end{figure}

Ion traps with integrated optical cavities are promising network nodes and have enabled e.g., photon-interfaced multi-qubit  processing registers \cite{repeater, shuttling}, near-deterministic photon extraction  \cite{JosefThesis, Stute2012, Stute2013}, evidence of strong ion-cavity coupling \cite{PhysRevLett.124.013602} and the entanglement of ions \SI{230}{\meter} apart \cite{NetworkPaper}. However, while multi-second qubit coherence times have been achieved in ion traps with lens-based photon collection \cite{PhysRevLett.130.090803}, such coherence times have not yet been reported in cavity-integrated ion traps. 
Qubit coherence time is a key networking parameter: it sets how long remote entanglement can be stored and should be much longer than the time required to generate more \cite{PhysRevA.89.022317, Hucul2015, Covey:2023qy}.  Achieving multi-second qubit coherence times in cavity-integrated ion-traps is non-trivial, as the cavity can introduce various decoherence mechanisms, including motional heating from the cavity mirror dielectric surfaces \cite{Ong_2020}, and fluctuating electric and optical fields associated with cavity length stabilization mechanisms.

In this work, we demonstrate multi-second coherence times for a register of qubits in a cavity-integrated trapped-ion network node. First, dynamical decoupling is used to extend the coherence times of a register of up to five ion qubits to the multi-second regime, without cavity-photon generation. Second, an ion-entangled cavity photon is generated from a single trapped-ion qubit, after which the ion-qubit is stored for up to two seconds, whilst still retaining strong entanglement. Third, the robustness of ion-qubit coherence times to the cavity-photon generation process is assessed by investigating how repeated attempts to obtain a photon from one ion in the register affect the quality of quantum information stored in co-trapped ions. Finally, calculations are presented showing that a moderately improved and duplicated version of the system should enable the simultaneous establishment of at least five Bell pairs between remote ion registers.

\section{Experimental setup and key methods} \label{main:setup}

Our experimental system consists of a string of $^{40}$Ca$^+$ atoms in a linear Paul trap with an integrated 20-mm-long Fabry-P\'erot optical cavity for photon collection at \SI{854}{\nano\meter} \cite{JosefPaper}.
The cavity axis is close to perpendicular to the ion string axis, with a relative angle of \SI{85.9}{\degree}.
The waist of the relevant vacuum mode of the cavity is \SI{12.31(8)}{\micro\meter} and the length of e.g., a five ion string with our chosen confinement is \SI{22(2)}{\micro\meter}, where the uncertainty is due to the uncertainty in our knowledge of the secular center of mass motional frequencies of the five ion string, given below. 
That geometry constrains only one ion in the string to be coupled to the cavity with the maximal coupling strength afforded by our cavity of $g_0 = 2\pi\times1.53$~\SI{}{\mega\hertz}~\cite{JosefThesis}; that ion must be positioned at a maximum in the \SI{854}{\nano\meter} cavity vacuum standing wave and center of the cavity waist. 
Figure~\ref{fig:intro} presents both an energy level diagram of $^{40}$Ca$^+$, showing e.g., the various qubit encodings used in this work, and a schematic of the ion-cavity setup. 
In cases where a single ion is trapped, the secular center of mass motional frequencies are $[0.9620(4),1.997(7),1.956(7)]$~MHz in order axial and two radials. In cases where five ions are trapped, the frequencies of those modes are changed to $[0.60(8),1.95(6),2.08(1)]$~MHz.
More details on the ion-cavity system are given in Appendix~\ref{appendix:experimental setup} and in \cite{JosefPaper}.

Ion-photon entanglement generation is realized via a bichromatic cavity-mediated Raman transition (BCMRT) \cite{Stute2012,JosefPaper}, driven via a single-ion-focused \SI{393}{\nano\meter} Raman laser beam with a \SI{1.2}{\mu\meter} intensity waist at the targeted ion \cite{repeater,101}. 
A Raman laser pulse on the ion %
ideally generates the maximally-entangled state $\ket{\psi}{=}(|D',H\rangle+|D,V\rangle)/\sqrt{2}$, where $|{D'}\rangle$ and $|{D}\rangle$ are the respective metastable Zeeman states $|3^2{D}_{5/2}, m_j = -3/2\rangle$ and $|3^2{D}_{5/2}, m_j = -5/2\rangle$ (Metastable qubit, Figure~\ref{fig:intro}a), and $|{V}\rangle$ and $|{H}\rangle$ are the vertical and horizontal polarization components of a single \SI{854}{\nano\meter} photon emitted into the cavity vacuum mode, respectively. After the generated photon has left the cavity, via the output mirror, it is coupled outside the vacuum chamber into a single mode optical fiber and sent to a polarization analysis setup (Appendix~\ref{appendix:experimental setup}). 
Two single mode optical fibers separately couple photons from the transmission and reflection ports of a polarizing beam splitter (PBS) in that setup, and send them to separate single-photon detectors. 
The cavity length is locked to resonance with injected \SI{806}{\nano\meter} laser light, via the Pound-Drever-Hall technique. To minimize energy-level shifts caused by this light, on an ion in the center of the cavity waist at \SI{854}{\nano\meter}, the cavity is locked to an \SI{806}{\nano\meter} spatial mode with an intensity minimum in the center (TEM01). When locked to a TEM00 mode, the \SI{806}{\nano\meter} mode has a waist of \SI{12.0(1)}{\mu\meter}.

The $|{D'}\rangle$ and $|{D}\rangle$ states, which encode our metastable qubit, decay via spontaneous emission with a lifetime of \SI{1.17}{\s}~\cite{James1998}. For longer term storage of quantum information, the ion qubit state is transferred into superpositions of the two stable ground states  $\ket{S}=|4^2{S}_{1/2}, m_j = -1/2\rangle$ and  $\ket{S'}=|4^2{S}_{1/2}, m_j = 1/2\rangle$, which serve to encode and store our memory qubit. Transfer of quantum information between the metastable- and ground-state qubit encodings is done via \SI{729}{\nano\meter} laser pulses. 
To protect the ground state qubit from decoherence due to e.g., ambient magnetic field fluctuations, we implement dynamical decoupling.
In particular, we implement the XY16 sequence, which protects arbitrary prepared qubit states against decoherence while mitigating the effect of imperfections in the pulse duration and frequency of the involved pulses~\cite{doi:10.1098/rsta.2011.0355, deLange:2010avk}. 
That sequence repeats 16 spin echos ($\pi$ pulses) with alternating axes of rotation, each spaced by a wait time of \SI{90}{\micro\second} (Appendix~\ref{appendix:pulsesequence}). Each spin echos lasts \SI{108(1)}{\micro\second} and is realized using a wire coil positioned outside the vacuum chamber, around which a current is driven at the ground state splitting frequency of \SI{11.6}{\mega\hertz}, set by our applied magnetic field of \SI{4.15}{\gauss} \cite{JohannesThesis}. The oscillating magnetic field generated by the wire coil couples equally to all co-trapped ions, leading to the same spin-echo time. 
All experiments presented in this work begin by Doppler cooling followed by optical pumping into the initial state $\ket{S}$.
To read out the logical ion-qubit states, we first transfer $\ket{S'}$ to $\ket{D'}$, thus converting from ground to optical qubit encodings. Then, the logical states are determined individually via fluorescence state detection, in which scattered \SI{397}{nm} photons are collected by an objective and imaged onto a digital camera ($\ket{S}=\ket{0}$ fluoresces, $\ket{D'}=\ket{1}$ does not).

\section{Multi-second memory-qubit coherence times}\label{sec:ionresults}

\begin{figure}[h!]
	\begin{center}
       \includegraphics[width=1\columnwidth]{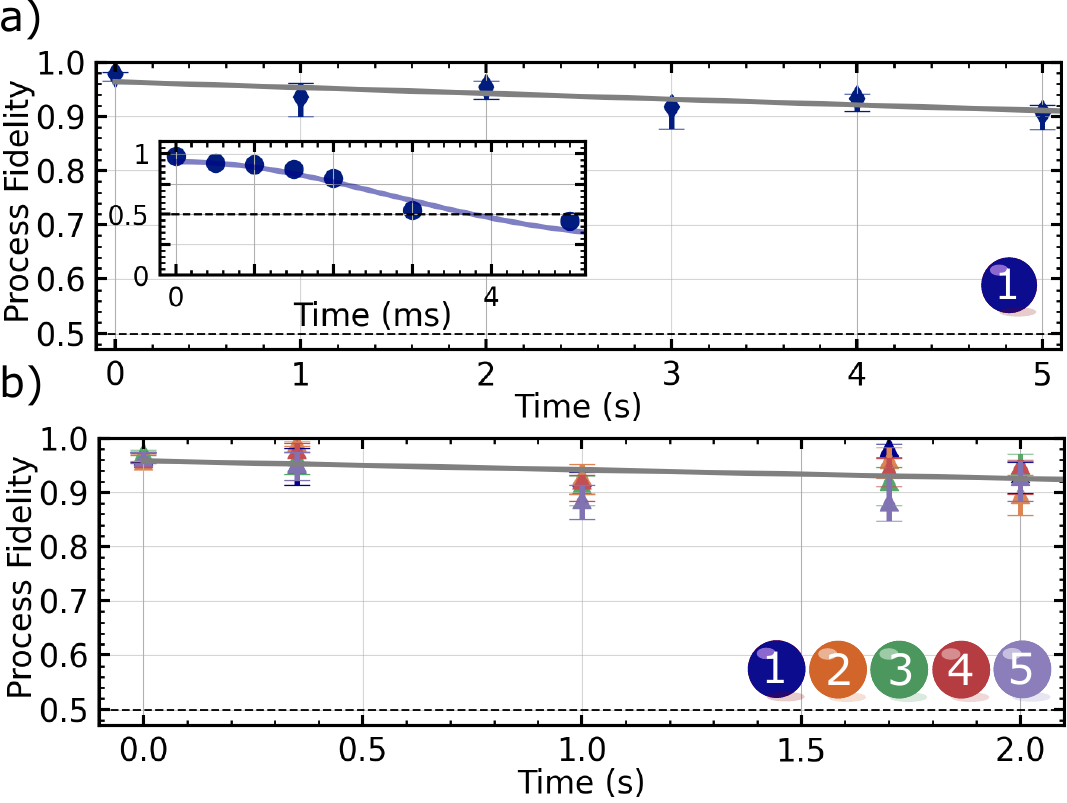}
		\caption{
   		\label{fig:ionmem} Memory qubit coherence enhanced by dynamical decoupling. Process fidelities with the ideal identity operator versus storage time for (a) one and (b) five co-trapped ions. Filled shapes: data; error bars represent one standard deviation of statistical uncertainty, derived from Monto Carlo analysis of tomographic data. Grey solid lines: fits to Eq.~\ref{eq:fid} in the main text. Black dashed lines show the 0.5 fidelity threshold for storage of remote entanglement. Inset (a): single-ion performance without spin echoes; blue line: model of \SI{50}{\hertz} magnetic-field noise.}
		\vspace{-5 mm}
	\end{center}
\end{figure}

\noindent\textbf{Single-ion coherence.} In the first experiment, we investigate the lifetime of the memory qubit when protected by the XY16 sequence, for a single ion and without cavity-photon generation.  %
The cavity remains unlocked, with no injected \SI{806}{\nano\meter} locking light. 
Each experimental run proceeds as follows:  the ground state qubit is prepared in a chosen input state, using the RF coil; the XY16 sequence is executed for a total time $t$; finally the ion qubit is measured in the chosen basis. Ion-qubit measurement in the logic basis proceeds as described in the previous section, with the modification that the fluorescence photons are collected by a photomultiplier tube (PMT), not the digital camera. Measurements in other bases are achieved by performing an additional RF pulse with a tunable phase before readout (Appendix~\ref{appendix:pulsesequence}). 
Experimental runs are repeated fifty times to enable reliable estimates of expectation values. 
Expectation values are estimated in the three Pauli basis $\sigma_z$ (logical), $\sigma_x$ and $\sigma_y$ for each of four input states: $\ket{0}$, $\ket{1}$, $\ket{+}=(\ket{0}+\ket{1})\sqrt{2}$ and $\ket{i}=(\ket{0}+i\ket{1})\sqrt{2}$.
The obtained data set is used to reconstruct the single-qubit process matrix $\chi(t)$ \cite{nielsen_chuang_2010} describing the memory evolution after time $t$. We calculate the fidelity $F$ between the reconstructed process matrix and the ideal identity process matrix $\mathcal{I}$, via the equation $F(\chi,\mathcal{I}) \equiv \operatorname{tr}\!\left( \sqrt{ \sqrt{\chi}\,\mathcal{I}\,\sqrt{\chi} } \right)^{2} $~\cite{PhysRevA.71.062310}. 
In Appendix~\ref{appendix:processfidelity} we show that $F>0.5$ serves as a threshold for the ground-state memory qubit's ability to successfully store remote entanglement: when that inequality is satisfied, the state fidelity of a stored ion-photon Bell state can remain above the threshold required for verifying entanglement. 
Estimated uncertainties in process fidelities are calculated via the Monte Carlo resampling technique.  
Our system control hardware allows for the implementation of dynamical decoupling sequences for up to five seconds when using the PMT, and up to two seconds when using the digital camera, limited by our current FPGA-based control system. %

The measured single-qubit process fidelities are found to remain well above the 0.5 threshold for all implemented memory-qubit storage times, with a minimum value of 0.91$^{+1}_{-3}$ for the maximum memory-qubit storage time of five seconds (Figure~\ref{fig:ionmem}a).  %
An analytical model of the form 
\begin{equation}\label{eq:fid}
	F(t) = Ae^{-t/\tau}+0.25
\end{equation}
is fit to the measured fidelities, where $A$ is a constant and $\tau$ is the memory-qubit coherence time. The fit yields $A=0.714(5)$ and $\tau=65(15)$s (Figure~\ref{fig:ionmem}a, gray curve). The model is derived from a single-qubit depolarization map (Appendix~\ref{appendix:imperfectmemroy}). We attribute the deviation from the ideal value $A=0.75$ to imperfections in state preparation. Repeating the experiment without dynamical decoupling yields a memory-qubit coherence time of a few milliseconds, which is well described by a model of the effect of ambient magnetic field fluctuations at \SI{50}{\hertz} with a random phase (Figure~\ref{fig:ionmem}a, inset). 
The amplitude of the \SI{50}{\hertz} magnetic field fluctuations used in the model is \SI{36(2)}{\micro\gauss}, which was measured in a separate calibration experiment. More details on that model are provided in Appendix~\ref{appendix:dephase}. %
We separately model the effect of those measured magnetic field fluctuations on the process fidelities $F$ achieved by the XY16 sequence, yielding a predicted sub-percent drop after two seconds of storage (Appendix~\ref{appendix:50hzxy16}). 
Remaining decoherence in our system, not mitigated by the spin echos, is attributed to the combined effect of fluctuations in the ambient magnetic field strength, and in the field strength generated by the RF coil at the ion, at frequencies higher than \SI{50}{\hertz}.

We find that the fitted memory-qubit coherence time $\tau$ with spin echos can vary from week to week by tens of seconds, for both single and multiple ions. The lowest recorded value to date was \SI{13(2)}{\second}. We attribute changing memory-qubit coherence times to changes in the ambient magnetic field noise in our building. 
 
 \noindent\textbf{Multi-ion coherence.} In the second experiment, we investigate the memory-qubit coherence times for five co-trapped ions.  
The experiment performed is identical to the previous one, except that the ion-readout is now done with the camera, not the PMT, such that individual ion states can be  determined.
The measured single-qubit process fidelities for all five ions remain well above threshold for the longest measured memory-qubit storage times of two seconds (Figure~\ref{fig:ionmem}b). There are no significant differences in memory performance across the ion string: for each measured time, the fidelity values for different ions are not more than two standard deviations apart. A fit to the depolarization model (Equation~\ref{eq:fid}), that considers data points from all ions, yields parameters $A=0.71(1)$ and a memory-qubit coherence time of $\tau=43(16)$s. 
We conclude that the multi-second coherence time of our single-qubit memory extends to five co-trapped memories. Specifically, each of the five co-trapped qubit memories perform as well as a single one, to within the precision of the available data. 

Our results here extend the previous state of the art for coherence times for individual ion-qubits that are first-order sensitive to magnetic field fluctuations \cite{Ruster2016}, in terms of maximum observed coherence time and ion-qubit number. For example, in \cite{Ruster2016}, a coherence time of up to 2.1(1)s was achieved for a qubit encoded into the spin ground states of a single $^{40}$Ca$^+$ ion using one spin echo.

\begin{figure}[t]
	\begin{center}
       \includegraphics[width=1\columnwidth]{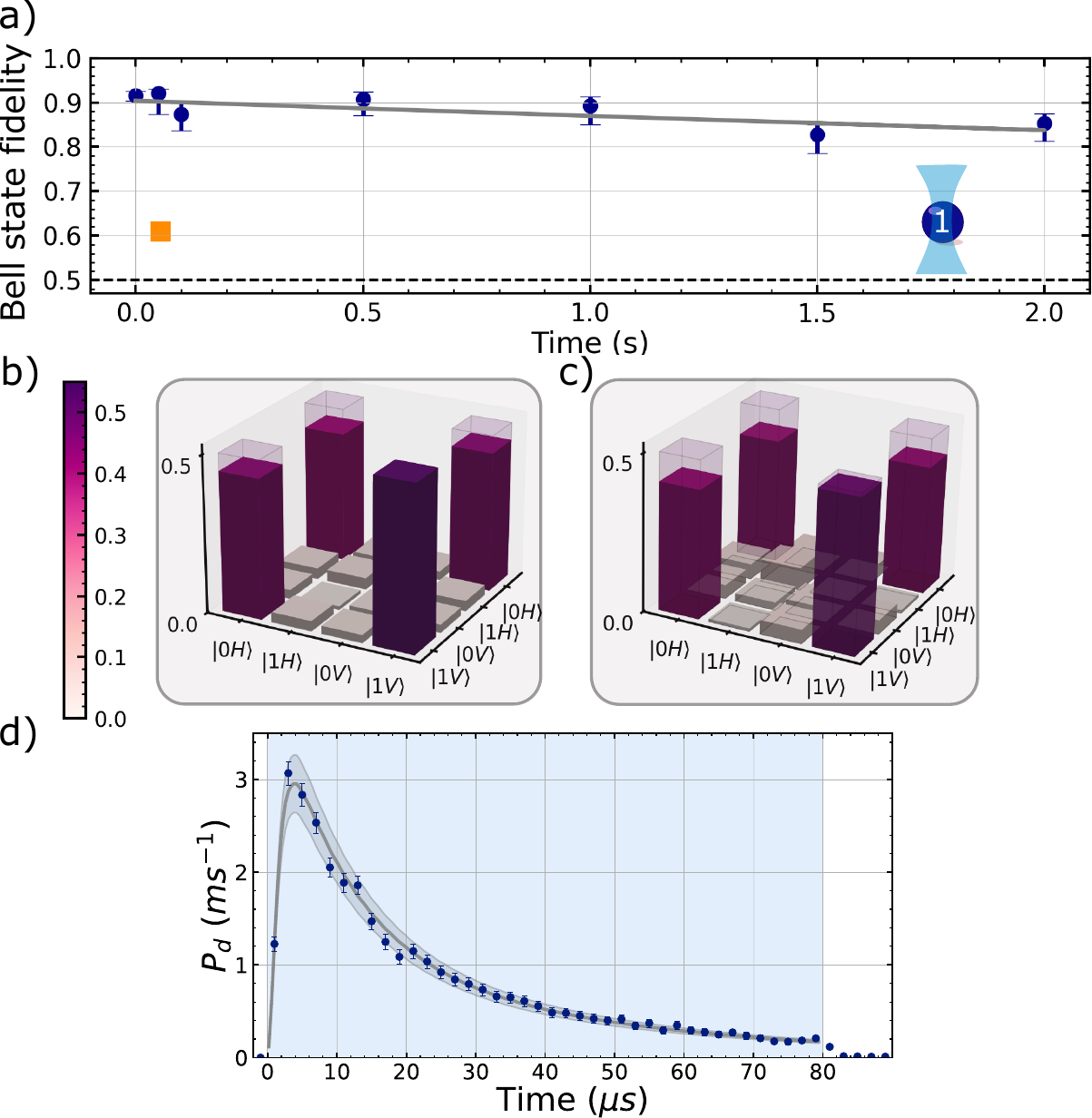}
		\caption{Storage of ion–photon entanglement. (a) Ion–photon state fidelity with the ideal Bell state versus storage time in a single trapped ion. Blue circles: data; error bars: one standard deviation. Gray solid line: fit to Eq.~\ref{eq:fid}. Orange square: result from Ref.~\cite{repeater}. Black dashed line at 0.5 indicates the entanglement threshold. (b),(c) Absolute values of tomographically reconstructed ion–photon density matrices at storage times of 0 s and 2 s, respectively. Transparent bars: ideal Bell state. (d) Histogram of the photon detection probability density $P_d$, averaged over all storage times. $P_d$ is defined as the number of detector counts, per \SI{2}{\micro\second} time bin, normalized by the number of photon generation attempts. Gray line: prediction from a master-equation model; shaded region: one standard deviation. Blue shaded region: \SI{80}{\micro\second} acceptance window used in the analysis. }	\label{fig:ionphoton}	
		\vspace{-5 mm}
	\end{center}
\end{figure}

\section{Storage of ion-photon entanglement}\label{ionphotonstoragesec}

We now turn our attention to using the long-lived memory states to store entanglement between an ion and a photon generated via the optical cavity. 
Thus, in the third experiment, a single ion is loaded into the trap and ideally placed at the optimal position for photon generation: a maximum of the \SI{854}{\nano\meter} vacuum standing wave at the center of the cavity waist. 
Each experimental run proceeds as follows: an \SI{80}{\micro\second} Raman laser pulse ideally generates the ion-photon entangled state $\ket{\psi}$ (defined in Section~\ref{main:setup}); in the case of photon detection, the metastable qubit is transferred to the ground state qubit; the XY16 sequence is executed for a total time $t$; ion-qubit readout in a chosen Pauli basis is done as described in the previous section, with the exception that the optional $\pi/2$ pulse is done optically using a \SI{729}{\nano\meter} laser (more details are provided in Appendix~\ref{appendix:pulsesequence}). 
The chosen ion-qubit and photon-qubit measurement bases are fixed throughout a single execution of the experimental pulse sequence. The sequence is repeated fifty times for each of the nine combinations of Pauli measurement bases for two qubits (ion and photon)—set by different combinations of waveplate settings for the photon qubit and laser pulses for the ion qubit—allowing for tomographic reconstruction of the two-qubit ion-photon density matrix $\rho(t)$. Laser light at \SI{806}{\nano\meter} is injected into the cavity for length stabilization at all times except during photon generation (Raman laser pulses), and during \SI{729}{\nano\meter} laser pulses, to eliminate AC Stark shifts on the metastable qubit. In cases in which no photon is detected, up to ten attempts to generate a photon are made. Ion-qubit readout is performed with the PMT and the maximum implemented wait time is two seconds. %

\noindent\textbf{Fidelity.} Ion-photon state quality is quantified by the Bell state fidelity $F_B(t)=Tr(U\rho(t)U^{\dagger}\ket{\psi}\bra{\psi})$, where $U$ is a one-qubit rotation of the photonic qubit that is applied numerically. %
The $U$ rotation employed is the one that yields a maximum fidelity for the measured ion-photon state with no wait time ($F_B(t=0)$), found by numerical search. This rotation does not change the entanglement content of the state and is necessary due to rotations suffered by the photon polarization qubit caused by the optical fiber between the vacuum chamber and the polarization analysis setup. A fidelity $F_B>0.5$ proves that the experimentally-reconstructed state $\rho(t)$ is entangled~\cite{Friis2019}. 
The measured Bell state fidelities (Figure~\ref{fig:ionphoton}a, blue data points) remain well above the threshold for proving entanglement for all ion-photon entanglement storage times, with a value after two seconds of $F_B(2s)=0.85^{+2}_{-4}$. 
The initial ion-photon Bell state fidelity $F_B(0)=0.92_{+1}^{-1}$ is lower than we have previously achieved in our system (0.966(5) in \cite{JosefPaper}). Identifying the origins of imperfections in this initial state is not the focus of this work.

We build a model for $F_B(t)$ based on single-qubit depolarization (of the ion qubit), as described in App.~\ref{appendix:imperfectmemroyionphoton}. Imperfections in the initial state of the ion-photon entangled state $\rho(0)$, are modelled via uncorrelated depolarization on both qubits, while temporal dynamics are modelled by depolarization of the ion qubit as before (Figure~\ref{fig:ionmem}). The model is found to yield the same analytical form for $F_B(t)$ as Equation~\ref{eq:fid}. Fitting that function to the data yields $\tau=20(4)$s and initial amplitude $A = 0.654(6)$ (Figure~\ref{fig:ionphoton}a, gray line).
By inserting those fitted parameters into Equation~\ref{eq:fid}, we calculate that $F_B(t)$ would drop below the 0.5 threshold for proving entanglement after \SI{18(3)}{\s}, with errors propagated from those of the two fitted parameters. We therefore expect that the ion-qubit memory should be able to store entanglement with the photon for up to \SI{18(3)}{\s}. 
The achieved ion-photon entanglement coherence time is two orders of magnitude greater than the best previous value of \SI{108(1)}{\milli\second} \cite{repeater} achieved in our system (Figure~\ref{fig:ionphoton}a, orange marker). %
Multi-second storage of ion-photon entanglement has previously been reported without an optical cavity and by encoding into a hyperfine transition in $^{43}$Ca$^+$ \cite{PhysRevLett.130.090803}, whose frequency is orders of magnitude less sensitive to magnetic fields than the relevant transition in the present work. 

\noindent\textbf{Photon efficiencies.} The probabilities with which an ion-entangled photon was generated and detected were measured to be $[6.76(8), 6.4(2), 6.0(2), 6.6(2), 6.1(2), 7.2(2), 7.3(3)]\times 10^{-2}$ for experiments implementing ion-photon entanglement with storage times of $[0, 0.05, 0.1, 0.5, 1, 1.5, 2]$~s, respectively. Uncertainties are calculated using Poissonian photon counting statistics. 
For the ion-photon entanglement storage time of \SI{0}{\micro\second}, the statistical uncertainty is smaller than for the other times due to a tenfold increase in the number of associated data points taken. 
Those photon detection efficiencies exhibit differences beyond those expected from Poissonian statistics.
Those differences are attributed to drifting optical-fiber coupling efficiencies in the polarization analysis setup, during the three hours over which data was taken for different ion-photon entanglement storage times. %

Figure~\ref{fig:ionphoton}d shows a histogram of all the single photon detection events measured over all seven implemented ion-photon entanglement storage times, revealing the single photon temporal wavepacket. The total photon detection efficiency is $0.0663(6)$, consistent with the average of the detection efficiencies for each individual storage time. 
The measured wavepackets associated with each ion-photon entanglement storage time all have a statistically consistent shape, and differ only by an overall reduction factor associated with the aforementioned drift in fiber coupling efficiency.

The measured wavepacket in Figure~\ref{fig:ionphoton}d is well described by a model of the photon generation process (Figure~\ref{fig:ionphoton}d, gray line), which predicts the wavepacket shape and detection efficiency. 
That model is based on a master equation comprising 18 electronic states in $\Ca$ and two frequency-degenerate modes of the cavity, details about which are provided in Appendix~\ref{app:wavepacketmodel}. 
Model parameters are set to values obtained from independent calibrations, except for two free parameters. 
The first free parameter is a reduction factor $\zeta$ in the coupling strength associated with the BCMRT, which we attribute to a displacement of the ion position from the maximum of the \SI{854}{\nano\meter} vacuum cavity standing wave. 
A value of $\zeta = 0.7$ was used, which provides a close match between the measured and modeled wavepackets shown in Figure~\ref{fig:ionphoton}d. That value of $\zeta$ would be consistent with an ion displacement of \SI{0.11}{\micro\meter} along the cavity axis, caused by improper positioning of the ion.  
The second free parameter is an efficiency reduction factor $P_{\mathrm{red}}$ associated with the aforementioned imperfectly aligned fiber coupling. 
The expected wavepacket efficiency is given by the product of the probability for the emission of a photon into the cavity predicted by the model ($0.47$), the independently characterised subsequent total detection path efficiency ($0.19(2)$) as detailed in Appendix~\ref{appendix:budget}, and a factor $P_{\mathrm{red}} = 0.75$. 
That product evaluates to 0.067(7), which is consistent with the measured wavepacket efficiency of $0.0663(6)$. %

Significantly higher total photon detection probabilities have been achieved in our system. For example, a total ion-entangled photon detection efficiency of 0.462(3) \cite{JosefPaper} was previously achieved by combining ground state cooling of the ion's motional state, an optimal drive laser direction with respect to the principal magnetic field axis, a lower-loss detection path and lower cavity scattering and absorption losses.

\begin{figure}[t!]
	\begin{center}
       \includegraphics[width=1\columnwidth]{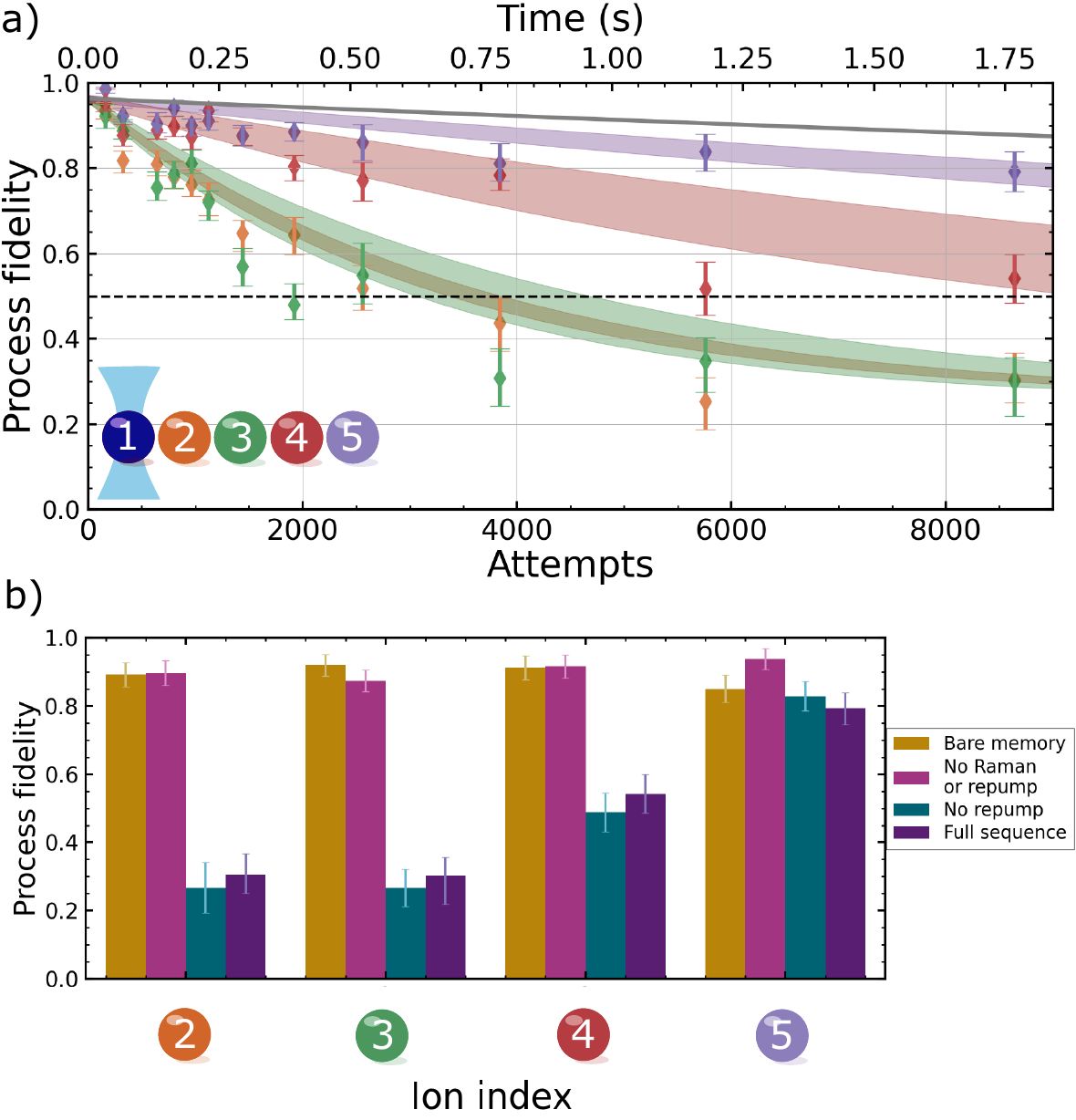}
\caption{Robustness of co-trapped ion-qubit memories. (a) Single-qubit process fidelities versus the number of cavity-photon generation attempts on ion 1. Color-filled symbols: data. Color-matched shaded regions: predictions from a model of Raman laser crosstalk; bounds indicate $\pm 1$ standard deviation. Gray line: independently calibrated bare-memory performance. Black dashed line at 0.5 indicates the threshold for survival of entanglement. (b) Process fidelities after 8640 attempts for four pulse sequences. Purple bars: full sequence (corresponding to 8640 attempts in panel (a)). Green bars: without repump lasers, preventing cavity photon generation. Pink bars: without Raman lasers, suppressing crosstalk; cavity-locking light remains. Yellow bars: bare-memory performance. Error bars: one standard deviation. %
}
\label{fig:robustness}
		\vspace{-5 mm}
	\end{center}
\end{figure}

\section{Memory Robustness}\label{robustnesssec}

 \noindent\textbf{Background.} The ability to robustly store quantum information in a network node, whilst attempting to establish new remote entanglement via the generation of photons is key to the realization of future quantum networks.
However, the process of generating photons from one ion in a register can compromise the quality of the quantum information stored in the states of co-trapped ions. 
It is thus important to assess how robust a memory qubit stored in one ion is to photon generation attempts on another co-trapped ion.  

Using ion traps with integrated lenses for photon collection (no cavities), there have been recent works investigating memory qubit  robustness. One approach co-traps ions of different species \cite{PhysRevLett.118.250502, PhysRevLett.130.090803}: an ion of one species produces networking photons while an ion of another serves as a spectrally-isolated quantum memory. This approach has been demonstrated for two ions \cite{PhysRevLett.118.250502, PhysRevLett.130.090803}, achieving unparalleled memory robustness at the expense of sub-optimal cooling and the need to reorder the ion string after background gas collisions using microfabricated traps. 
Another approach uses a register of ions of the same species and exploits different qubit encodings \cite{10.1063/5.0069544} to realize different networking functionalities in a way that mitigates unwanted crosstalk. Using integrated lenses for photon collection, that single-species approach has been investigated for e.g., two Yb${^{+}}$ ions \cite{Yang:2022il, Feng:2024fy}, and for two \cite{PhysRevLett.134.070801} and for five \cite{Cui_2025} $^{40}$Ca$^+$ ions. 

In this final series of experiments, we present results from an investigation into the memory robustness in our cavity-integrated single-species architecture. Specifically, we investigate how qubits stored in the spin ground states of four $^{40}$Ca$^+$ `memory' ions tolerate repeated cavity photon generation attempts on a fifth co-trapped $^{40}$Ca$^+$ `communication' ion, as shown schematically in Figure~\ref{fig:intro}b. %
In general, the designation of communication and memory ion need not be permanent as each ion can serve either function (communication or memory), as recently shown in our system by shuttling ions in and out of the cavity waist \cite{shuttling}. 
Memory robustness has previously been investigated for up to two ions in our system \cite{repeater}, revealing no statistically significant degradation of memory performance for up to 200 cavity-photon generation attempts on the neighboring ion. Here, we employ five ions and up to 8640 cavity-photon generation attempts which is sufficient to clearly resolve the current capabilities of, and limits to, memory robustness.  

There are many potential sources of memory ion decoherence introduced by cavity-photon generation. For example, the memory ions are exposed to the \SI{806}{\nano\meter} laser light used for cavity-length stabilization, \SI{854}{\nano\meter} cavity photons themselves, spontaneously emitted \SI{393}{\nano\meter} photons from the communication ion as well as Raman laser crosstalk. 
By Raman laser crosstalk, we refer to the presence of residual laser intensity on ions neighboring the targeted ion. %

\noindent\textbf{Experimental details.}
The communication ion (ion 1) at the end of the ion string is placed at the optimal position in the cavity for photon generation, as previously described and shown conceptually in Figure~\ref{fig:intro}b.
Process tomography of the ground state ion qubits of all four co-trapped ions is carried out as described in Sec.~\ref{sec:ionresults}, except that a composite X5 sequence~\cite{ultrahighfidelity} is now used to transfer the qubit state $\ket{S'}$ of each ion to $\ket{D'}$ before readout, instead of a single $\pi$ pulse. 
The composite X5 sequence is implemented via five sequential \SI{729}{\nano\meter} $\pi$ pulses with different phases (Appendix~\ref{appendix:pulsesequence}), aiming in the end to perform a $\pi$ rotation that is more robust to the temperature of the ion string than a single $\pi$ pulse. 
During the time gap between the involved state preparation and readout pulses, $N$ photon generation attempts (Raman pulses) are applied to the communication ion, interleaved with spin echos applied to the entire string. Specifically, each attempt consists of a Raman pulse on the communication ion (\SI{70}{\mu\s}), an RF spin echo (\SI{108}{\mu\s}) followed by repump laser pulses at \SI{866}{\nano\meter} and \SI{854}{\nano\meter} to empty the metastable manifolds (\SI{5}{\mu\s}). 
The spin echos are separated by \SI{90}{\micro\second}, as in the previous experiments. 
The reconstructed process matrices of the four memory qubits are used to calculate the process fidelity with the ideal identity as before. 

In contrast to the ion-photon entanglement experiment reported in Figure~\ref{fig:ionphoton}, three changes were made to the photon generation and characterization processes. First, readout of the ion qubits is now performed at the end of every experimental run, independent of whether a photon was detected in any attempt. Second, the polarization state of the generated photons is not characterized; photons are sent directly to a detector.  The first and second changes increase the data accumulation rate for ion-qubits, allowing for a more precise characterization of the memory states. Third, the implemented Raman pulses are now monochromatic, which ideally generate the product ion-photon state $\ket{D,V}$. That choice was made as the focus here is on the performance of the co-trapped memories, not the photon generation.

\subsection{Results}

The performance of the four memory ions is first characterized in the absence of photon generation attempts and with an unlocked cavity (the `bare' memory performance). 
That characterization is a repeat of the experiment presented in Figure~\ref{fig:ionmem}b, but now taken on the day on which the subsequent experiments described here are performed.
The repeated characterization is performed to account for a change in the memory-qubit coherence times between the days on which the two sets of experiments were done, which we attribute to a change in the magnetic field noise environment experienced by the ions.
A fit of Equation~\ref{eq:fid} to all obtained process fidelities yields a memory-qubit coherence time of $\tau=13(2)$s and initial amplitude $A=0.713(5)$. 
The full protocol is now implemented, including photon generation attempts with a locked cavity, and process matrices are reconstructed as a function of the number of photon generation attempts made.
Comparison between the measured memory fidelities (Figure~\ref{fig:robustness}a, colored markers) and the fitted bare memory performance (Figure~\ref{fig:robustness}a, top gray line) reveals a marked reduction in the presence of photon generation. The general trend is that the fidelities of memory ions closer to the communication ion decay faster than those of ions further away.  The numbers of attempts for which the measured fidelities remain over two standard deviations above the 0.5 threshold required for sustaining memory ion-photon entanglement, are $[1920, 1120, 3840, 8640]$ for ions 2-5, respectively.

\noindent\textbf{Model of laser crosstalk.}
We develop a model for memory ion decoherence caused by Raman laser crosstalk. 
Although the laser is focused onto the communication ion, one expects a finite intensity on the co-trapped ions. That intensity can off-resonantly excite the memory ions from the ground state to the excited $\Pp$ manifold which has a $\tau_P = $~\SI{6.9}{\nano\second} lifetime~\cite{PhysRevLett.70.3213}, decaying via spontaneous emission. Such spontaneous emission will destroy phase coherence in the memory qubit state. The rate of off-resonant excitation is proportional to $\gamma^{\mathrm{eff}}_i = \gamma(\Omega_i/2\Delta)^2$, where $\Omega_i$ is the Rabi frequency of the laser coupling to ion $i$, $\gamma = 1/\tau_P $ is the spontaneous scattering rate of the $\Pp$ manifold, and $\Delta = $~\SI{400}{\mega\hertz} is the laser detuning from the $\Ss\leftrightarrow\Pp$ transition. 
We use the aforementioned 18-level master equation model to predict the effect of off-resonant excitation on the single-qubit process fidelities encoded in the memory ions, as described in Appendix~\ref{appendix:fullmodel}. 
The model includes the `bare' depolarization experienced by the memory ions when protected by dynamical decoupling (Equation~\ref{eq:fid}) parameterized by $\tau$. The predicted process fidelity decay is well described by the function 
\begin{equation}
\label{crosstalk}
F_i(t) = Ae^{-t/\tau}e^{-0.67\gamma^{\mathrm{eff}}_iB t}+0.25. 
\end{equation}
where $B$ is the fraction of the total time $t$ during which the Raman laser is on, which for the present experiment has the value $B=0.35$. 
The square of the Rabi frequency $\Omega_i^2$ is proportional to the laser intensity at ion $i$ and to the scattering rate $\gamma^{\mathrm{eff}}_i $. 
The values $\Omega_i^2$ are measured in a calibration experiment performed shortly before the data in Figure~\ref{fig:robustness}a was taken (Appendix~\ref{appendix:adderr}).
The obtained relative intensities are $\Omega_i^2/\Omega_1^2=$[26(2),24(5),7(3),1.9(5)] ppm, for ions 2-5 respectively, where $\Omega_1^2 = \num{211(5)e1}$~\SI{}{\mega\hertz\squared}. 
Using the aforementioned value for the initial amplitude $A$, the model for Raman laser crosstalk now has no free parameters. The model predictions (Figure~\ref{fig:robustness}a, colored lines) show clear qualitative agreement with the data.  Clearly Raman laser crosstalk is the limiting factor in co-trapped memory qubit lifetimes in the presence of photon generation.

\noindent\textbf{Other error sources.}
To determine if the detrimental effects of other error sources can be resolved, the full experiment is repeated for 8640 photon generation attempts,  
turning off different potential contributions one at a time.
The purple bars in Figure~\ref{fig:robustness}b show the same data as for the full experiment with 8640 photon generation attempts in Figure~\ref{fig:robustness}a. %
First, we only turn off the repump laser pulses at \SI{866}{\nano\meter} and \SI{854}{\nano\meter} (Figure~\ref{fig:robustness}b, green bars) and find that this has no statistically significant effect on the memory fidelities. 
In the absence of those repumpers, photons are essentially no longer generated from the communication ion after the first attempt, as the electron is trapped in the metastable $D_{5/2}$ manifold. Therefore, neither \SI{854}{\nano\meter}  cavity photons, nor spontaneously emitted \SI{393}{\nano\meter} photons, generated by the communication ion over 8640 Raman pulses have a resolved detrimental effect on co-trapped memory qubits. 
Next, in addition to the repump laser pulses, the \SI{393}{\nano\meter} Raman laser pulses are also turned off (Figure~\ref{fig:robustness}b, pink bars), leaving only the \SI{806}{\nano\meter} cavity locking laser light, and find that the memory fidelities recover to the same values as the bare memory (Figure~\ref{fig:robustness}b, yellow bars). 
The \SI{806}{\nano\meter} cavity locking laser light therefore has no measured effect on memory performance.  
We thus conclude that, to within the statistical precision afforded by our data, Raman laser crosstalk is entirely responsible for the degradation of the co-trapped ion-memories. 

\noindent\textbf{Photon efficiencies.}
The single photon detection efficiencies obtained for the results in Figure~\ref{fig:robustness}a are now described. Over the 8640 attempts to generate a photon from the communication ion, an average of 345(1) photons were detected, corresponding to a 3.99(1)\% average detection efficiency. Over the first 100 attempts, the average detection efficiency was higher at 5.7(1)\%. The efficiency drop is caused by the scattering of  \SI{393}{\nano\meter} photons from the communication ion heating the ion string up. As the ion string heats, the coupling strength of the photon generation process is reduced by e.g., coupling of the Raman laser to motional sidebands of the ion string \cite{shuttling}. 

Further analysis of the temporal shape of the detected single photon wavepacket, for the results shown in Figure~\ref{fig:robustness}a, is presented in Appendix~\ref{app:wavepacketmodel}. We estimate that, when taking the data in Figure~\ref{fig:robustness}, 
the cavity-photon generation process was erroneously driven approximately \SI{0.24}{\mega\hertz} off-resonant. For comparison, the cavity-mediated Raman transition was measured to have a full-width at half-maximum of \SI{0.60(4)}{\mega\hertz}. 
A model of the monochromatic photon generation process predicts that resonant driving would yield a significantly higher 9.5(9)\% detection efficiency, for the initial ion-string temperature. We do not expect changes in the ion string temperature over the first few 100 attempts to have a significant effect on the photon generation probability. 
We subsequently repeated the first 100 attempts for an on-resonant drive and obtained a detection efficiency of 10.0(1)\%, consistent with the model's prediction.  Moreover, for the very first photon generation attempt, on-resonant driving achieved a 13.7(3)\% photon detection efficiency. This can be understood as follows: before the first attempt, the communication ion is optically pumped into the ideal initial state $\ket{S}$, yielding the highest expected photon generation efficiency. The absence of optical pumping before subsequent attempts means that the communication ion is in a statistical mixture of $\ket{S}$ and $\ket{S'}$, yielding a lower photon generation efficiency. Optical pumping of the communication ion after each attempt could be performed in future after e.g., temporarily moving memory qubits to the $\Dd$ manifold or by exploiting a single-ion focused \SI{729}{\nano\meter} laser.  Such `addressed' optical pumping has previously been demonstrated in $^{40}$Ca$^{+}$ strings e.g., \cite{Kirchmair:2009kh}. 

Maintaining photon generation efficiency for arbitrary attempt number $N$ should be possible in future by combining optical pumping with periodic cooling of the ion string back to the initial Doppler temperatures. Such cooling could be done by adding additional $^{40}$Ca$^{+}$ ions to the string on which to perform Doppler laser cooling at \SI{397}{\nano\meter}, whilst ground-state memory qubits encoded in other ions are temporarily moved into metastable qubit states---states that are hidden from the Doppler cooling process. Such `in-sequence' cooling in single-species ($^{40}$Ca$^{+}$) strings has been successfully performed before e.g., \cite{Kirchmair:2009kh, doi:10.1126/science.aad9480}.

\noindent\textbf{Prospects for cross-talk minimisation.} %
The investigation clearly shows that the memory performance is currently limited by Raman laser crosstalk during cavity-photon generation attempts.   
The measured relative intensities  $\Omega_i^2/\Omega_1^2$ are compatible with the results of simulations of the optical path of the Raman laser~\cite{MarcoThesis}: laser crosstalk on co-trapped ions is limited by Airy fringes in the focal plane due to the finite numerical aperture of the employed focusing objective (NA $= 0.3$). 
There are various approaches to reducing the intensity of Raman laser crosstalk in future. First, moving to an objective with a higher numerical aperture would enable a reduction of the amplitude of Airy fringes.  Second, ions could be positioned into Airy fringe minima by adjusting the axial trapping potential. While microfabricated traps with segmented electrodes could in principle provide sufficient electric potential control to position ions individually, our current non-segmented harmonic trap cannot. Third, using a different cavity with a reduced mode volume (and therefore higher ion-photon coupling strength) would allow the photon generation rate to be maintained for a reduced overall laser intensity. 
Alternatively, the memory ions could be shuttled away from the communication ion once they contain qubits storing remote entanglement. Indeed, such architectures, that exploit segmented ion traps to shuttle ions between communication and memory/computation zones have been proposed e.g., \cite{Monroe_2013}.  The present results show that moving memory ions at least \SI{20}{\micro\meter} away from a communication ion is already sufficient to almost entirely mitigate crosstalk, in our single-species architecture.

\section{Towards entangling remote ion-qubits registers}\label{outlooksec}

The results from the previous section show that co-trapped memory qubits should be capable of protecting quantum information while thousands of attempts to make new remote entanglement via cavity photon generation take place. In this section, we present a calculation that shows that the current memory performance allows for the simultaneous establishment of five and more entangled remote-ion pairs between two duplicates of the ion-cavity system. 

Consider duplicating our ion-cavity system, yielding two remote multi-qubit registers with identical performance. Each system contains at least five co-trapped ions, labelled $i=1:5$ (c.f., Figure~\ref{fig:intro}b). Entanglement is established between remote ion-pairs ($i$-$i$) via the established two-photon detection scheme~\cite{RevModPhys.82.1209}, which was previously used to entangle an ion in our system with an ion in a cavity-integrated trap in another building \cite{NetworkPaper}. Specifically, each system emits a single ion-entangled photon and, conditional on detection of two photons after a beamsplitter, the ions are projected into a remote Bell state. 
The probability of establishing remote ion entanglement per attempt is $P_c^2/2$, where $P_c$ is the probability that a cavity photon generation attempt on the communication ion in either system leads to a single photon detection event at any of the detectors (when considering the four-detector scheme). The average number of attempts required to establish entanglement between a remote ion-pair is thus $\langle N_{\mathrm{av}} \rangle=2/P_c^2$.  
We use the conservative value of $P_c=0.1$, which was already achieved for the first 100 attempts for resonant driving in this work. 
That value could be maintained with periodic re-cooling using additional co-trapped $^{40}$Ca$^{+}$ ions, as previously described. 
The average number of attempts on the communication ion in each system to establish a remote Bell pair is then $\langle N_{\mathrm{av}} \rangle=200$.

We define the link efficiency $\xi_i=N_{\mathrm{max},i}/ \langle N_{\mathrm{av}} \rangle$, where $N_{\mathrm{max},i}$ is the maximum number of attempts on the communication ion ($i=1$) in each system for which a pre-established Bell pair stored between remote memory ions ($i=2:5$) remains entangled.
Imagine that four remote-ion Bell pairs have already been established and are each stored in a different remote pair of ion memory qubits (matched $i$ value from 2 to 5). Attempts are now made on the communication ions to establish a new remote-ion Bell pair. $\xi_i$ quantifies the average number of times a new Bell pair could be established between communication ions before the Bell pair stored between the $i^{th}$ memory ion pair is lost. Achieving a link efficiency over one represents a key threshold to surpass for emerging quantum network architectures \cite{PhysRevA.89.022317, Hucul2015, Covey:2023qy}. 

As further described in the Appendix~\ref{appendix:fidelityremotebell}, we use the measured process matrices of our four memory ions (used to produce the data in Figure~\ref{fig:robustness}a) to calculate the largest attempt number for which the fidelity of a stored ideal remote ion-ion Bell state would still be above the threshold value of 0.5 for proving entanglement. The results yield $N_{\mathrm{max},i}= [800, 960, 1920, 8640]$ and corresponding link efficiencies of $\xi_i=[4,4.8,12.8,43]$, for memory ions 2-5 respectively. Therefore, the communication ions could be entangled up to 4 times on average before the first pair of remote memory ions would lose their entanglement. The last remote memory pair to lose entanglement would do so after on average 43 new Bell pairs are made between the communication ions.   %

The predicted performance should be more than sufficient to establish all five remote Bell pairs, beginning with none. The experiment could proceed as follows: remote entanglement is made between the communication ions and stored in the ground state manifolds of those ions; the ion string in each system is translated such that the neighboring ion is now positioned in the center of the cavity \cite{shuttling} and remote entanglement is established and mapped to the ground state; this continues until all ions are remotely entangled, interspersed with spin echos and re-cooling when needed. 
Using this procedure, after each successful remote entanglement event the ion strings are translated relative to the cavity. Over time, the established entangled pairs are moved progressively farther from the Raman-beam foci, thereby reducing decoherence from laser crosstalk. The first remotely entangled pair must survive all subsequent photon-generation attempts and therefore is expected to have the lowest final fidelity.
The appropriate $N_{\mathrm{max},i}$ for that ion pair, accounting for the movement of the ion string, will be approximately the average of $[800, 960, 1920, 8640]$, which is $3080$. 
The number of attempts required in order to establish all four other Bell pairs with a probability of 0.99 is $2006$ (as determined from the negative binomial distribution with success probability $P^2_c/2$). Therefore, the calculation predicts that all five remote Bell pairs can be sequentially established and the entanglement in them can survive until the end of the procedure. 

Key next experimental steps are to achieve in-sequence optical pumping and cooling, in order to stabilise the photon generation process over repeated attempts. On the other hand, a probability of $P_c=0.462(3)$ for the detection of an ion-entangled photon has previously been achieved in our system \cite{JosefPaper}. If that performance were duplicated in two systems, the total number of attempts  required to establish five remote Bell pairs with 0.99 probability is below 100. In that case, in-sequence cooling may not be required.  \\

\section{Conclusion}\label{conclusion}

Multi-second coherence times were demonstrated in a cavity-integrated register of trapped-ion qubits. This performance was enabled by storing qubits in long-lived ground-state spin-memory states that are spectrally well separated from the cavity-stabilization laser, together with an RF-driven dynamical-decoupling sequence that is robust to pulse errors and ion-string temperature fluctuations. These extended coherence times enabled storage of ion-photon entanglement for up to two seconds with only a few percent reduction in Bell-state fidelity, representing a substantial improvement over the previous state of the art in cavity-integrated ion-trap systems.

The robustness of the ion-qubit memories under repeated attempts to generate cavity photons from a co-trapped ion was investigated. After thousands of attempts, memory degradation was found to be caused primarily by laser crosstalk, and several mitigation strategies were proposed. In the future, the demonstrated memory performance could already enable the generation of many entangled ion pairs between two remote cavity-integrated ion registers. To achieve this, it would be beneficial to combine the methods presented here with in-sequence cooling and optical pumping, which have been demonstrated elsewhere.

Ion registers can each serve as a high-fidelity universal quantum processor, and our results present a promising route for efficiently connecting the Hilbert spaces of remote register copies, in line with a modular approach to scalable quantum computing \cite{PhysRevA.89.022317}. In the near term, the generation of multiple remote Bell pairs between two registers could provide a testbed for investigating e.g.,: the teleportation of multi-qubit states between registers \cite{PhysRevLett.70.1895, PreskillPH229-Ch3}; the purification of entanglement for efficient long-distance quantum information transfer within a repeater architecture \cite{PhysRevLett.81.5932}; blind quantum computing~\cite{5438603, PhysRevLett.132.150604, 10.1007/s00224-018-9872-3}, and methods for efficient determination of fidelities between the states of remote multi‑qubit registers via a non‑local SWAP test \cite{PhysRevLett.88.217901, Gottesman:1999oe,  PhysRevA.62.052317}. The long-coherence times achieved in this work are one of the key ingredients for the long-distance cavity-integrated quantum repeater chains envisioned in \cite{repeater} (ion-ion) and \cite{hvsx-cx2d} (ion-hybrid).

Datasets will be provided on reasonable request.

\begin{acknowledgments}
The Austrian Science Fund (FWF) 
[Grant DOIs: 10.55776/PAT1502824, %
10.55776/F71, %
10.55776/P34055 %
 and 10.55776/COE1]; %
the European Union under the DIGITAL-2021-QCI-01 Digital European Program under Project number No 101091642 and project name `QCI-CAT', and the European Union’s Horizon Europe research and innovation programme under grant agreement No. 101102140 and project name ‘QIA-Phase 1'; the Österreichische Nationalstiftung für Forschung, Technologie und Entwicklung (AQUnet project).  
B.P.L. acknowledges support from the CIFAR Quantum Information Science Program of Canada. The opinions expressed in this document reflect only the author’s view and reflects in no way the European Commission’s opinions. The European Commission is not responsible for any use that may be made of the information it contains. For open access purposes, the author has applied a CC BY public copyright license to any author-accepted manuscript version arising from this submission. 

J.B. took the experimental data. 
J.B., J.H., M.C., V.Kru., and B.P.L. developed the experimental setup.
J.B. and  B.P.L. carried out the data analysis and interpretation. 
The modeling was done by J.B. 
The manuscript was written by B.P.L. and J.B. and all authors provided detailed comments. 
The project was conceived and supervised by B.P.L.
\end{acknowledgments}

\appendix

\section{Experimental setup}\label{appendix:experimental setup}

The cavity parameters used for simulations in this work are the same as those given in~\cite{shuttling}. 
Specifically, our near-concentric Fabry{\textendash}Perot optical cavity is formed by two mirrors, each with a radius of curvature \SI{9.9841(7)}{\milli\meter}, separated by a length \SI{19.906(3)}{\milli\meter}.  
The finesse of the cavity at \SI{854}{\nano\meter} was measured to be $30(1)\times 10^3$.  
The cavity decay, measured by ringdown, is $\kappa = 2\pi \times 126(2)$~\SI{}{\kilo\hertz}. The quantity $\kappa$ is equal to the half-width at half-maximum of the cavity resonant spectral linewidth. 
From the amplitude of the locked-cavity error signal we estimate a cavity-lock jitter of \SI{9(2)}{\kilo\hertz} \cite{JosefThesis}. For details about the construction and initial characterization of cavity see Ref.~\cite{JosefThesis}, and for its use to achieve near-deterministic photon extraction see Ref.~\cite{JosefPaper}. 

For analysis of the polarization of the \SI{854}{\nano\meter} cavity photons, they are sent via a single-mode fiber to a polarization analysis setup. 
In that setup, the photons are first sent through a motorized half and quarter waveplate, followed by a polarizing beam splitter (PBS). 
Finally, photons at the two output ports of the PBS are each coupled into a separate single mode optical fiber, and sent to separate superconducting nanowire single-photon detectors. 
Those detectors are measured to have efficiencies of 0.88(3) and 0.80(3) at \SI{854}{\nano\meter}, and an average measured dark count rate of 0.4~cps (measured with a dark cap on the fiber to the detector). Values are taken from the Supplemental Material of Ref.~\cite{shuttling}). 

\section{Pulse sequences}\label{appendix:pulsesequence}

This section first introduces some notation used throughout. It then presents the pulse sequence implemented in the experiment investigating the memory robustness of a five‑ion string (Section~\ref{robustnesssec}). Finally, it details the modifications to that sequence used in the experiments described in Sections~\ref{sec:ionresults} and~\ref{ionphotonstoragesec} of the main text.

We introduce the ideal single-qubit rotation
\begin{equation}\label{eq:rotglob}
    R(\theta, \phi) = e^{-i\frac{\theta}{2}({\sigma_x}\cos{\phi} + {\sigma_y}\sin{\phi})},
\end{equation}
where $\theta$ and $\phi$ are real values corresponding to the rotation angle and phase, respectively. 
These ideal single-qubit rotations are implemented on all ion-qubits in two different ways: using the RF coil, which rotates ground state qubits encoded in the states $\ket{S}$ and $\ket{S'}$, and using a \SI{729}{\nano\meter} laser, which rotates optical qubits encoded in a state in the $\Ss$ manifold and another state in the $\Dd$ manifold. 
The duration of a $\pi$ pulse, $R(\pi, \phi)$, using the RF coil is \SI{108(1)}{\micro\second}, and using the \SI{729}{\nano\meter} laser is between \SI{5}{\micro\second} and \SI{10}{\micro\second} for all transitions used. The duration of $\pi/2$ pulses is approximately half of the aforementioned values. Those pulse durations are valid for all of the $R(\theta, \phi)$ rotations presented in the rest of this section. 

The pulse sequence used for the experiment investigating the memory robustness of a five-ion string (Section~\ref{robustnesssec}) is now provided, as an enumerated list of steps. The durations of specific steps are given in parentheses.  
\begin{enumerate}
	\item \SI{393}{\nano\meter} laser pulse applied for intensity stabilization. After transmitting through the vacuum chamber, the pulse is captured by a photodiode and used as part of a sample and hold system to return the intensity of subsequent pulses to a target set point. (\SI{1}{\milli\second})
	\item Doppler cooling using the \SI{397}{\nano\meter} laser and repumping lasers at \SI{854}{\nano\meter}  and \SI{866}{\nano\meter} (The duration of this step varied depending on the storage time, as described after this list). 
	\item Optical pumping to the state $\ket{S}$ (defined in Figure~\ref{fig:intro}) using a $\sigma_-$ polarized \SI{397}{\nano\meter} laser (\SI{25}{\micro\second}).
	\item The \SI{806}{\nano\meter} laser light, that is otherwise sent into the optical cavity at all times for length stabilization is switched off (few microseconds).
	\item Preparation of the initial state of the ground state ion-qubits, for state tomography. If the desired input state is $\ket{S}$, no operations are performed. If the desired input state is $(\ket{S} + e^{i\phi}\ket{S'})/\sqrt{2}$, an RF $\pi/2$ pulse realizing the rotation $R(\pi/2, \phi + \pi/2)$ is applied. If the desired input state is $\ket{S'}$, an RF $\pi$ pulse is applied. A total of four different initial states are prepared, which are defined in the main text. 
	\item The \SI{806}{\nano\meter} laser is switched back on, sending it into the cavity for its length stabilization (few microseconds)
	\item Start of a loop, with iteration indexed by the integer $n$
	\item Pulses of repumping lasers at \SI{854}{\nano\meter} and \SI{866}{\nano\meter} empty the $\Dd$ and $\Ddt$ manifolds (\SI{5}{\micro\second})
	\item The \SI{806}{\nano\meter} laser light, that is otherwise sent into the optical cavity at all times for length stabilization is switched off (few microseconds).
	\item Monochromatic \SI{393}{\nano\meter} Raman pulse (photon generation) on Ion 1 (\SI{70}{\micro\second})
	\item The \SI{806}{\nano\meter} laser is switched back on, sending it into the cavity for its length stabilization (few microseconds).
	\item Application of an RF $\pi$ pulse realizing the rotation $R(\pi, \Phi_{n|16})$. Here, $\Phi_{n|16}$ is the $\mathrm{mod}_{16}(n)$-th element of the list of phases in Equation~\ref{eq:XY16phases}, where the modulus operator $\mathrm{mod}_{16}(n)$ corresponds to the remainder after dividing $n$ by 16. 
	\item End of $n^{\mathrm{th}}$ iteration of loop: go back to start of loop (Step 7) until $n = N_l$ loops are completed. $N_l$ is chosen to implement the desired storage time.  
	\item The \SI{806}{\nano\meter} cavity-locking light is again switched off (few microseconds).
	\item Preparation of the tomographic measurement basis of the ground state ion-qubits. If the desired measurement basis is $\sigma_z$, no operations are performed. If the desired measurement basis is $\sigma_x$ ($\sigma_y$), an RF $\pi/2$ pulse is applied on the $\ket{S}\leftrightarrow\ket{S'}$ transition, realizing the rotation $R(\pi/2, \pi/2)$ ($R(\pi/2, 0)$). 
	\item The repumping pulses of step 8 are repeated (\SI{5}{\micro\second})
	\item A composite X5 sequence is implemented by five sequential \SI{729}{\nano\meter} $\pi$ pulses applied on the $\ket{S'}\leftrightarrow\ket{D'}$ transition. Those pulses realize rotations $\prod_{k = 1}^6R(\pi, \phi_k)$, where the five phases represented by $\phi_k$ follow the elements of the list in Equation~\ref{eq:X5phases}. 
	\item Fluorescence state detection of the ion-qubit states using \SI{397}{\nano\meter} laser and repumping lasers at and \SI{866}{\nano\meter}. Ion fluorescence is collected by a camera. (\SI{5}{\milli\second})
\end{enumerate}

The total combined duration of steps 8-11, including delays for switching on and off relevant acousto-optic modulators, amounts to \SI{90}{\micro\second}, consistent with the \SI{90}{\micro\second} wait time between RF spin echos in the experiments described below.
The duration of Doppler cooling in Step 2 ($t_{\mathrm{Dop}}$) is increased in proportion to the storage time $t$, approximately following the relation $\Delta t_{\mathrm{Dop}}\approx0.4\Delta t$, where $\Delta t_{\mathrm{Dop}}$ and $\Delta t$ signify a change in the associated quantity.
For the smallest implemented storage time of $t = 32$~\SI{}{\milli\second}, the Doppler cooling time is $t_{\mathrm{Dop}} = $~\SI{13}{\milli\second}. 
Increasing $t_{\mathrm{Dop}}$ in that way is found to be necessary to allow our FPGA-based experimental hardware to run the above sequence for storage times up to $t=5$~s  for a single ion. 

The phases of the composite $\pi$ pulse in Step 15 are given by
\begin{equation}\label{eq:X5phases}
	\Phi_{X5} = \{0.0672\pi, 0.3854\pi, 1.1364\pi, 0.3854\pi, 0.0672\pi\},
\end{equation}
which implement the X5 sequence derived in~\cite{ultrahighfidelity}. 
The phases of the spin echo $\pi$ pulse in Step 10 follow the list
\begin{align} 
\Phi_{XY16} = &\{0, \pi/2, 0, \pi/2, \pi/2, 0, \pi/2, 0,\label{eq:XY16phases}\\ &\pi, 3\pi/2, \pi, 3\pi/2, 3\pi/2, \pi, 3\pi/2, \pi\}, \nonumber
\end{align}
which implement the XY16 dynamical decoupling sequence~\cite{doi:10.1098/rsta.2011.0355}. 
The total number of loops $N_l$ is chosen such that ion-qubits are stored for the desired time. For example, a storage time of approximately two seconds is realized for $N_l = 640$.

We now describe how the above sequence is modified for the experiments investigating the coherence time of a register of five trapped ion-qubits in the absence of cavity-photon generation (Section~\ref{sec:ionresults}). 
That same modified sequence is also used for investigating the coherence time of a single trapped ion-qubit, with the fluorescence collected by a photomultiplier tube instead of a camera. 
The \SI{806}{\nano\meter} light for stabilizing the cavity length is not used. 
Steps 8-11 associated with photon generation are replaced with a wait time of equal combined duration (\SI{90}{\micro\second}). 
The composite $\pi$ pulse in step 15 is replaced with a single $\pi$ pulse, whose performance is better in the absence of photon generation. 
Steps 1 and 14 are removed as they are not necessary in the absence of photon generation attempts. 
The tomographic rotation preparing the desired input state in step 5 is performed using the \SI{729}{\nano\meter} laser on the $\ket{S}\leftrightarrow\ket{D'}$ transition. A $\pi$ pulse %
is then applied on the $\ket{S'}\leftrightarrow\ket{D'}$ transition. %
There is no important reason for exchanging the RF pulse in the original step 5 with those  \SI{729}{\nano\meter} pulses; they ideally prepare the same initial qubit state in the ground state. 
The tomographic rotation preparing the desired measurement basis in step 13 is analogously performed using the \SI{729}{\nano\meter} laser on the $\ket{S}\leftrightarrow\ket{D'}$ transition, and is preceded by a $\pi$ pulse on the $\ket{S'}\leftrightarrow\ket{D'}$ transition.

Next, we describe the sequence used for the experiment investigating ion-photon entanglement storage using a single ion (Section~\ref{ionphotonstoragesec}), which involves a different set of modifications to the sequence with steps 1:16. 

Step 5 is replaced by a second loop (Loop 2), with iteration indexed by the integer $m$. Each iteration consists of one attempt. 
An attempt consists of \SI{80}{\mu\second} of repumping, followed by \SI{160}{\mu\second} optical pumping, followed by an \SI{80}{\mu\second} bichromatic Raman laser pulse, which ideally generates the ion-photon entangled Bell state $\ket{\psi}$ defined in the main text. 
In the case in which a photon is successfully detected, Loop 2 is terminated and 
two $\pi$ pulses are then applied, one on the $\ket{S}\leftrightarrow\ket{D}$ and the other on the $\ket{S'}\leftrightarrow\ket{D'}$ transition, which map the metastable qubit to the ground state. The sequence then moves to step 6.
In the case in which no photon is detected after an attempt in Loop 2, that loop is iterated up to ten times ($m\leq 10$). In the case in which no photons are detected in all iterations of Loop 2, the entire sequence restarts. 
The tomographic rotation preparing the desired measurement basis in step 13 is now performed using the \SI{729}{\nano\meter} laser, following the description in the previous paragraph.
Steps 14 and 15 are removed, because in this case the qubit is already encoded in the bright $\ket{S}$ and the dark $\ket{D}$ levels, so the mapping is not needed before fluorescence detection.

\section{Photon path efficiency budget}\label{appendix:budget}

A detailed efficiency budget list for the single photon detection path is now given, based on independent measurements. The following list is specifically for the experiment in which a single ion stored entanglement with a cavity photon,  presented in Section~\ref{ionphotonstoragesec}.  The beginning of each element in the following list gives the transmission probability associated
with a distinct part of the detection path, beginning with losses experienced after the photon is emitted into the
cavity and ending in photon detection. The product of the list of efficiencies quantifies the total photon path efficiency $P_{\mathrm{path}}$, used in photon wavepacket simulations described in Sec.~\ref{app:wavepacketmodel}.

\begin{enumerate}
	\item 0.43(3): Probability that a photon in the cavity subsequently exits the cavity through the output mirror and into the desired output mode. 
	\item 0.96(1): Transmission of free-space optical elements between cavity output mirror and first fiber coupler (see $P_{el}$ in~\cite{JosefPaper}).
	\item 0.84(3): Coupling efficiency into the first single mode fiber, at the output of the cavity. The value was measured with \SI{854}{\nano\meter} laser light. 
	\item 0.79: Fiber joiner. This efficiency is the lower bound according to the datasheet provided by the manufacturer.
	\item 0.82(3) and 0.82(3): Efficiencies of the two paths in the polarization analysis board (transmission and reflection respectively) ending with fiber-coupled photons at the two output ports of the analysis board, measured with \SI{854}{\nano\meter} laser light. 
	\item 0.80(3) and 0.88(3): Efficiencies of the two detectors at \SI{854}{\nano\meter} (transmission and reflection paths, respectively), calibrated by comparison with a manufacturer-calibrated detector. 
\end{enumerate}

The products of the above efficiencies are 0.18(2) and 0.20(2) for the transmission and reflection detection paths, respectively, corresponding to an average value of $P_{\mathrm{path}}=0.19(2)$ for the ion-photon entanglement experiment. 
We consider $P_{\mathrm{path}}$ an upperbound for the achieved detection path efficiency for the experiments where ion-photon entanglement is generated, because all values in the list above were independently optimized when characterized, and the combined efficiency is expected to be only lower. 

For the experiment in which a monochromatic Raman laser pulse triggers emission of a single cavity photon from a five-ion chain (Section~\ref{robustnesssec}) a modified path efficiency is used for simulations. The photon is sent straight to a single detector, bypassing the polarization analysis board (element five in the list above is therefore excluded). 
For element six, the detector with efficiency 0.80(3) was used. 
The resulting modified path efficiency is $P_{\mathrm{path}}=0.22(2)$.

\section{Calibration of Raman laser crosstalk}\label{appendix:adderr}

Section~\ref{robustnesssec} provides independently measured values of the crosstalk of the Raman laser on each co-trapped `memory' ion, when the laser is focused onto ion 1. 
Those values are measured using a Ramsey experiment to estimate the size of the AC Stark shift imparted by the Raman laser on the $\ket{S}$ state of each of those memory ions.
The measured AC Stark shift for each ion is then used to calculate the Rabi frequency $\Omega_i$ of the Raman laser coupling to the $\Ss\leftrightarrow\Pp$ transition of each memory ion $i$. 
Here, we provide a detailed sequence for the experiment used to obtain those values. 

The measurement begins with five co-trapped ions, under the same trapping conditions used for the experiment reported in Section~\ref{robustnesssec}. After Doppler cooling and optically pumping each ion into the state $\ket{S}$, a \SI{729}{\nano\meter} $\pi$/2 pulse is applied on the $\ket{S}\leftrightarrow\ket{D'}$ transition (as defined in Figure~\ref{fig:intro}a). The Raman laser is then applied to the communication ion for \SI{20}{\milli\second}. A $\pi$-pulse is applied on the same transition, followed by a wait time of 20ms. A second $\pi$/2 pulse is then applied on that transition, with a phase whose value is scanned in sequential experiments. A final readout in the logical basis via fluorescence detection captured by the camera recovers Ramsey oscillations for each ion-qubit separately. The total phase of the Ramsey oscillations for each memory ion is compared to a repeat of the experiment in which the application of the Raman laser is replaced with a wait time of equal duration. 
The phase shift between the two Ramsey oscillations for ion $i$ ($\phi_i$) is expected to be $\phi_i = \delta^{AC}_iT$, where $T = $~\SI{20}{\milli\second} is the duration for which the Raman laser is applied and $\delta^{AC}_i$ is the AC Stark shift imparted by the Raman laser on memory ion $i$. 
That relation is used to make an estimate of $\delta^{AC}_i$ for each memory ion from the measured $\phi_i$ values. 
For the polarization and propagation direction of our Raman laser with respect to our magnetic field, the AC Stark shift is expected to be of magnitude $\delta_i^{AC} = \Omega_i^2/6\Delta$. 
That relation is used to make an estimate of $\Omega_i^2$ for each memory ion from the measured $\delta_i^{AC}$ values. 

The much larger Rabi frequency of the Raman laser coupling to the targeted communication ion is measured in a different way. Specifically, we measure the AC Stark shift it induces on the resonance frequency of the cavity-mediated Raman transition, as described in e.g., Appendix B3 in Ref.~\cite{JosefPaper}.

\section{Memory performance thresholds for entanglement storage}

This section introduces two performance thresholds for the ability of a memory qubit to faithfully preserve the entanglement contained in a Bell pair.  
The first threshold considers one half of a Bell pair to be stored by the memory qubit, with the other half stored in an ideal memory. 
In this case, a qubit memory with a process matrix $\chi$ is found to successfully store the entanglement of the Bell pair when the threshold $F(\chi,\mathcal{I}){>}0.5$ is fulfilled. 
The second, more stringent threshold considers both qubits of a Bell pair to be stored by memory qubits that are both described by the same process matrix $\chi$. 
In this case, the fidelity ($F_B$) of the output Bell state after storage is calculated with the initial, ideal Bell state. 
The memory qubit is above threshold when $F_B$ is above the threshold value certifying the existence of entanglement: $F_B > 0.5$~\cite{Friis2019}. 

\subsection{One imperfect memory}\label{appendix:processfidelity}

In Section~\ref{sec:ionresults} it was stated that $F(\chi,\mathcal{I}){>}0.5$ is an important performance threshold for a single memory qubit, where $F(\chi,\mathcal{I})$ is the process fidelity for some storage time $t$ (Fig. \ref{fig:ionmem}). 
The precise definition of this threshold is now given, followed by a short proof.  

\emph{Definition:} Consider that a memory qubit is used to store one half of a Bell pair.  Initially the state is perfect. The other half is encoded in some remote system which we here assume undergoes no decoherence (perfect memory). If the process fidelity of the memory qubit for storage time $t$ is above 0.5 ($F>0.5$) then entanglement still exists in the stored state.

\emph{Proof:} The value of this threshold can be immediately obtained from the Choi-Jamio\l kowski isomorphism~\cite{CHOI1975285, JAMIOLKOWSKI1972275}. From that isomorphism is holds that $F(\chi,\mathcal{I})= \bra{\phi}\rho\ket{\phi}$, where $\ket{\phi}$ is the Bell state 
\begin{equation}\label{app:eq:bell}
	\ket{\phi} = (\ket{00} + \ket{11})/\sqrt{2}
\end{equation}
and $\rho$ is the state after application of the $\chi$ matrix to one qubit in $\ket{\phi}$ ~\cite{PhysRevA.71.062310}. The quantity $\bra{\phi}\rho\ket{\phi}$ is directly the Bell state fidelity  $F_B$ between the perfect Bell state and the one stored in the qubit memory. The state fidelity threshold for entanglement in the stored state is $F_B>0.5$ ~\cite{Friis2019}, completing the proof.

\subsection{Two imperfect memories}\label{appendix:fidelityremotebell}

\begin{figure}[t!]
	\begin{center}
       \includegraphics[width=1\columnwidth]{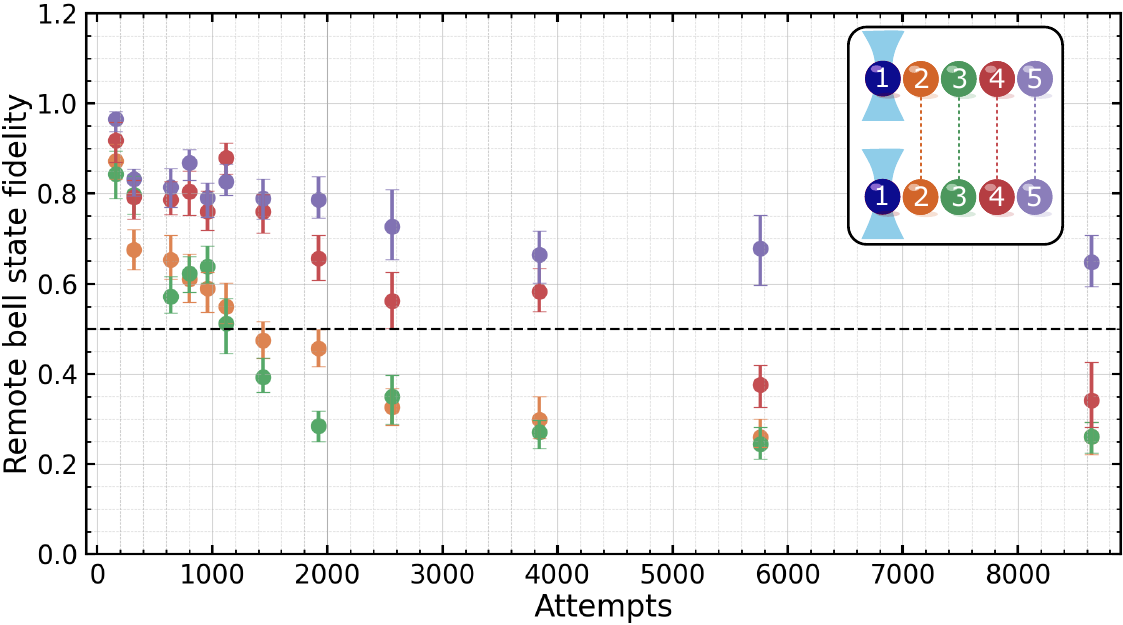}
		\caption{Predicted remote ion-ion Bell state fidelity decay, as a function of attempts to generate a new remote Bell pair. Inset panel: remote memory-ion pairs store Bell pairs, while attempts to make new remote Bell pairs are made (laser pulses). Main panel: color-filled shapes show predictions,  color-matched to ions in the inset.  Data is color-coded by ion. Black dashed line: entanglement threshold. Error bars are one standard deviation of statistical uncertainty }\label{fig:appremoteent}
		\vspace{-5 mm}
	\end{center}
\end{figure}

In our outlook (Section~\ref{outlooksec}), we are interested in estimating when an initially perfect Bell state stored in two remote ion-memory qubits would lose entanglement. 
The imperfect action of both memory qubits now degrades entanglement over time. 
Even in the case in which the process matrices of both memories are known, we do not know how to derive an analytical process fidelity threshold for entanglement survival in this scenario. 
Instead we take experimentally measured memory process matrices, apply them to a perfect Bell state, and determine if the output is above the Bell state fidelity threshold for proving entanglement ($F_B>0.5$)~\cite{Friis2019}. That procedure is now described in more detail.

We start with a perfect Bell pair $\ket{\phi_{ii}}$ stored between two remote ion-memory qubits with matched string position index $i$. Specifically, we consider two remote 5-qubit ion strings, where ion 1 in each string is the communication ion, in the focus of the optical cavity and ions $i=2:5$ are memory ions in order of distance from ion 1. Next, the same single-qubit process matrix $\chi(N)_{i}$ is applied to both qubits in $\ket{\phi_{ii}}$, where $N$ is the number of photon generation attempts on ion 1 in each string. 
The process matrices $\chi(N)_{i}$ are the ones obtained from the experiment investigating memory robustness in the five-ion chain (Figure~\ref{fig:robustness}a). Finally, we calculate the remote Bell state fidelity of the output state. The obtained fidelities, for each remote memory ion pair, are shown in Figure ~\ref{fig:appremoteent} as a function of attempt number $N$. The predicted $N_{\mathrm{max},i}$ values correspond to the implemented attempt number for which that fidelity, for ion-pair $ii$ first becomes within two standard deviations of the threshold value of 0.5 (Figure~\ref{fig:appremoteent}). 
Those values are $N_{\mathrm{max},i}=[800,960,1920,8640]$ for ions 2-5, respectively: these are the predicted attempt numbers for which an initially perfect remote Bell state stored between remote ion-memory qubits would still be entangled.

\section{Master equation model of photon wavepackets}\label{app:wavepacketmodel}

A master equation model is numerically solved to obtain predictions for the temporal wavepackets of single photons emitted into the optical cavity. 
Predicted wavepackets are obtained for both ion-photon entangled states, shown in Figure~\ref{fig:ionphoton}d, and unentangled photons shown in Figure \ref{fig:appphoton}. This section first provides general information about the master equation model. The subsequent subsections give detailed information about the two aforementioned cases considered.

The master equation model includes all relevant eighteen Zeeman levels contained within the $\Ss$, $\Pp$, $\Ppo$, $\Dd$ and $\Ddt$ manifolds of a single $\Ca$ ion, as well as two orthognal polarization modes of the optical cavity. 
A detailed summary of the master equation model, coded in Python, is given in Chapter 3 of~\cite{DarioThesis}.  
The model initializes the ion in the state $\ket{S}\bra{S}$, and finds the ion-cavity state $\rho(t)$ after the \SI{393}{\nano\meter} laser has been applied to the ion for a time $t$. 
For a fixed set of model parameters, the single photon temporal wavepacket, $P_d(t)$, is calculated via the equation $P_d(t) = P_{\mathrm{path}}2\kappa\mathrm{Tr}[\rho(t)(a^{\dag}a + b^{\dag}b)]$, where the path efficiency $P_{\mathrm{path}}$ is introduced in Appendix~\ref{appendix:budget}, the cavity decay $\kappa$ is introduced in Appendix~\ref{appendix:experimental setup}, $a^{\dag}$ ($a$) is the raising (lowering) operator of the horizontal cavity mode, and $b^{\dag}$ ($b$) is the raising (lowering) operator of the vertical cavity mode. The total probability that a photon is detected within a time window of length $t_{R}$ is calculated via the equation $P_{\mathrm{det}} = \int_0^{t_{R}}dtP_d(t)$.

\subsection{Simulation of ion-entangled photon wavepackets generated by a bichromatic Raman process}\label{app:simbichro}

How we obtain the simulated wavepacket shown in Figure~\ref{fig:ionphoton}d is now described.  The model parameters include the experimental geometries of the Raman laser, of the cavity and of the magnetic field, as well as the jitter of the resonance frequency of the cavity, which are the same as described in~\cite{repeater} (see Section I.B and III.C of the Supplemental Material).  The other key model parameters are now presented.\\

\noindent \emph{Ion-cavity coupling strength---} The ion-cavity coupling strength was set to a value of $g = \zeta g_0$, where maximal coupling strength $g_0$ is defined in the main text, and $\zeta$ is a reduction factor between $0$ and $1$ which accounts for any reduction in the achieved ion-cavity coupling. 
We use a value of $\zeta = 0.7$, which provides a close match between the measured and modeled wavepackets. 
We attribute this reduction factor to a displacement of the ion's position away from the maximum of the \SI{854}{\nano\meter} cavity standing wave.

\noindent \emph{Rabi frequencies---} The bichromatic Raman laser has two tones with Rabi frequencies $\Omega_1$ and $\Omega_2$, which ideally generate a cavity photon with $V$ and $H$ polarization, respectively. 
In order to estimate the values of those Rabi frequencies, we measure the AC Stark shift imparted by the bichromatic Raman laser on the $\ket{S}$ state of a trapped ion, as described in Appendix I.B of Ref.~\cite{repeater}. 
We measure the AC Stark shift to be \SI{0.58(5)}{\mega\hertz}. 
The procedure used to extract the Rabi frequency estimates from that AC Stark shift measurement is provided in Appendix III.C of Ref.~\cite{repeater}, resulting in values used for the simulations of $\Omega_1 = $~\SI{23.3}{\mega\hertz} and $\Omega_2 = $~\SI{27.5}{\mega\hertz}. 
The coupling strength of the cavity photon generation process produced by either tone is proportional to $\zeta g_0 \Omega^r_i$. Here, the reduced Rabi frequency $\Omega^r_i=\alpha \Omega_i$, where $\alpha$ is a reduction factor ($0\leq \alpha \leq 1$) and $ \Omega_i$ is the Rabi frequency of tone $i$. The reduction factor accounts for unwanted coupling of the Raman laser to the thermal motion of the ion in the radial direction (direction of the laser). The reduction factor is a function of the phonon number in each of the two radial motional modes and thus, for a finite motional temperature, is a statistical distribution. 

We use a Monte Carlo method to simulate the effect of the radial motion on the laser-ion coupling, as detailed in Appendix~II.C of Ref.~\cite{shuttling}. %
The two radial modes of the single ion after Doppler cooling are each expected to follow a thermal distribution, each with a mean phonon number that is separately measured to be $n_\mathrm{av} = 7.3(4)$, corresponding to a temperature $T = 0.75(4)$~\SI{}{\milli\kelvin}. 
That number was obtained by measuring Rabi flops on a \SI{729}{\nano\meter} transition of the single ion and fitting them to a simple model within the Lamb-Dicke approximation. 
For the Monte Carlo simulation, $200$ samples of the two parameterized thermal distributions are randomly made, each yielding two phonon numbers. For each sample, the laser-ion reduction factor $\alpha$ is calculated via Equation~S1 in Ref.~\cite{shuttling}. 
The master equation is then solved, yielding a sample photon wavepacket $P_d(t)$, where a path efficiency value of $P_{\mathrm{path}} = 0.19$ is used (see Appendix~\ref{appendix:budget} for the corresponding efficiency budget). 
Those $200$ sampled photon wavepackets are finally averaged to yield a single simulated wavepacket $\tilde{P}_d(t)$ that accounts for thermal motion. 
The total probability that a photon is detected within a time window of length $t_{R} = $~\SI{80}{\micro\second} (Raman laser pulse length) is calculated via $\tilde{P}_{\mathrm{det}} = \int_0^{t_{R}}dt\tilde{P}_d(t) = 0.089(9)$. 

\noindent \emph{Detection path efficiency correction---} As discussed in Section~\ref{ionphotonstoragesec}, the amplitude of the measured photon wavepacket (markers in Figure~\ref{fig:ionphoton}d) is lower than $\tilde{P}_d(t)$, which we attribute to imperfectly aligned fiber coupling in the detection path. 
We account for this by multiplying $\tilde{P}_d(t)$ by $P_{\mathrm{red}} = P_{\mathrm{av}}/\tilde{P}_{\mathrm{det}} = 0.75$, where $P_{\mathrm{av}} = 0.0663(6)$ is the measured detection probability averaged over the seven implemented storage times shown in Figure~\ref{fig:ionphoton}d. 
The simulated wavepacket shown in Figure~\ref{fig:ionphoton}d is $P_{\mathrm{red}}\tilde{P}_d(t)$, with the grey area representing one standard deviation in measurement uncertainty in $P_{\mathrm{path}}$ (Appendix~\ref{appendix:budget}).

\subsection{Simulation of polarised photon wavepackets generated by a monochromatic Raman process}

Section~\ref{robustnesssec} presented our investigation into the robustness of our four co-trapped ion memories to cavity photon generation from ion 1. There, we make use of a model of the monochromatic cavity photon generation process.  
That model is based on the bichromatic model from the previous subsection with the following three adaptations: the amplitude of the second laser tone is set to zero; the initial state of the ion before each cavity photon generation is modified to account for the absence of optical pumping; and a different detection path.  In the following, key parameters are presented, followed by a description of optical pumping adaptation and finally model predictions are compared to experimental data.

The same value of the ion-cavity coupling reduction factor $\zeta = 0.7$ is used as for the model in Appendix~\ref{app:simbichro}. 
The single tone of the Raman laser ($\Omega_1$) was measured to implement an AC Stark shift of \SI{1.47(3)}{\mega\hertz} on the $\ket{S}$ state of ion 1. 
That Stark shift leads to a value $\Omega_1 = $~\SI{60}{\mega\hertz} used in the model. 
A \SI{0.24}{\mega\hertz} detuning of the Raman laser from resonance is used in the simulation, which is found to provide a close match between the measured and modeled wavepackets. %
 A path efficiency of $P_{\mathrm{path}}=0.22(2)$ is used (Appendix~\ref{appendix:budget}). 
The path efficiency reduction factor is set to unity ($P_{\mathrm{red}} = 1$): the imperfect fiber-coupler is now absent as photons are sent straight to a detector, avoiding the polarization analysis board.
The Monte Carlo method to account for laser coupling to motion is used as before, but now each sample contains ten phonon numbers: one drawn from each of ten radial modes of motion of the five ion string. 
Those samples are randomly drawn from the associated thermal distributions, each set to the same temperature $T$ measured for a single ion (Appendix~\ref{app:simbichro}). 

As described in Section~\ref{robustnesssec}, the absence of optical pumping before each attempt results in ion 1 being in a statistical mixture of $\ket{S}$ and $\ket{S'}$ before all attempts except the first. 
To investigate the effect of that lack of optical pumping, a separate set of wavepacket measurements were made after each of the first 16 photon generation attempts, this time with more data points. 
That investigation revealed that for all attempts after the first two, the ion-qubit is well described by the initial ion state $\rho(0) = 0.42\ket{S}\bra{S} + 0.58\ket{S'}\bra{S'}$ in the model, which we use for the initial ion state in the master equation model. 

\begin{figure}[t!]
	\begin{center}
       \includegraphics[width=1\columnwidth]{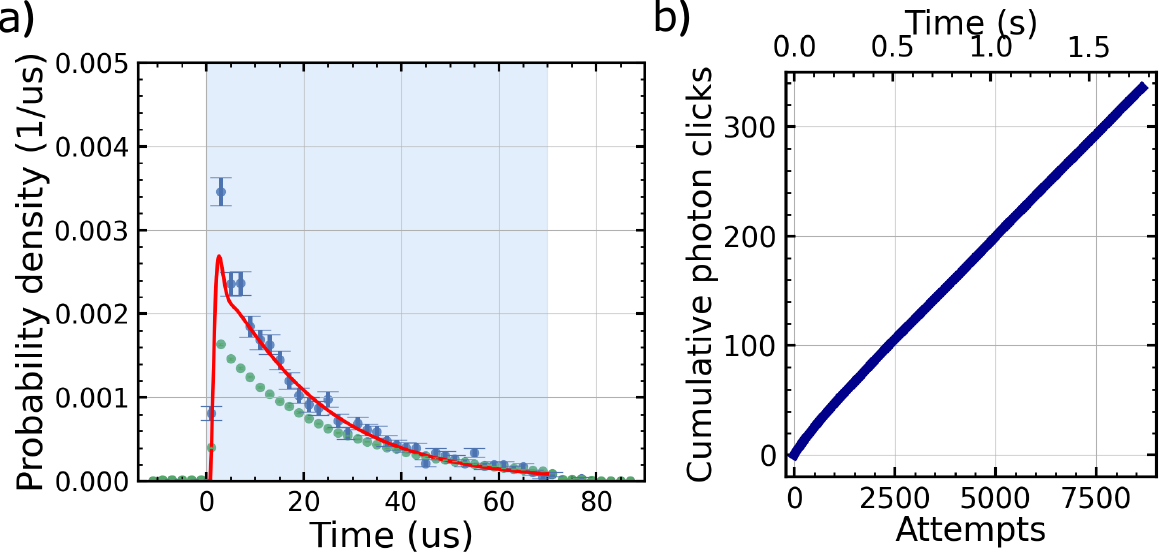}
		\caption{Detected photons during repeated photon generation attempts on ion 1, in the five-ion chain. Data is taken from the same experiment that produced the results in Figure~\ref{fig:robustness}. \textbf{(a)} 
		Histograms of the photon detection probability density $P_d$. $P_d$ is defined as the number of detector counts, per \SI{2}{\micro\second} time bin, normalized by the number of photon generation attempts. Blue markers are the photon wavepacket averaged over the first 100 photon generation attempts. Detection efficiency is determined within the \SI{70}{\micro\second}-long acceptance window shown by the blue shaded area. Green markers are the photon wavepacket averaged over all 8640 photon generation attempts for comparison. Error bars are one standard deviation of uncertainty, calculated from Poissonian statistics and in some cases smaller than marker size. Red line is a simulation from the monochromatic model described in the Appendix. \textbf{(b)} Cumulative measured photon clicks registered within the acceptance window.} \label{fig:appphoton}
	\end{center}
\end{figure}

Simulated photon wavepackets are now compared to the data recorded in the memory robustness investigation. Figure~\ref{fig:appphoton}a presents a comparison of the measured photon wavepacket over the first 100 attempts on ion 1 (blue shapes) and the corresponding simulation (red line). 
The simulation predicts a detection efficiency of $P_{\mathrm{det}} = 5.7(5)\%$, where the error is propagated from that of $P_{\mathrm{path}}$, which is consistent with the measured value of 5.7(1)\% (integration of blue shapes over the blue time window in Figure~\ref{fig:appphoton}a ). 
The measured photon wavepacket over all 8640 attempts (Figure~\ref{fig:appphoton}a, green shapes) achieves a lower efficiency of  3.99(1)\%, which we attribute to ion-string heating. Figure~\ref{fig:appphoton}b shows the cumulative photon clicks achieved over the 8640 photon generation attempts.

\section{Modelling decoherence in the memory ions}

This section contains details on all the models developed to simulate decoherence of the ion-qubit memories. These models are used to predict the memory-qubit process fidelities  obtained in the different experiments presented in the main paper.

\subsection{Effect of 50 Hz noise on a memory qubit unprotected by dynamical decoupling}\label{appendix:dephase}

In Section~\ref{sec:ionresults} an experiment without dynamical decoupling yields a memory-qubit coherence time of a few milliseconds, that is well
described by a model of the effect of ambient magnetic field fluctuations at 50 Hz with a random phase (Figure~\ref{fig:ionmem}a, inset). Here we give details on that model. 

The model considers a magnetic field of the form 
\begin{equation}\label{50hzb}
	B(t) = B_0\sin{(2\pi 50t + \beta)},
\end{equation}
where $B_0$ represents the amplitude of the oscillation, $t$ represents time, and $\beta$ is a phase term. 
That magnetic field interacts with the ground-state qubit via a linear Zeeman Hamiltonian of the form 
\begin{equation}\label{linzee}
	H = \hbar g\mu_BB(t)\sigma_z/2,
\end{equation}
where $g = 2$ is the Land\'e factor for the $\Ss$ manifold, and $\mu_B$ is the Bohr magneton.
The phase ($\Phi$) generated by the off-diagonal terms of the ion-qubit density matrix after being exposed to that magnetic field for a time $t$ is 
\begin{align}
    \Phi &= \frac{\mu_BgB_0}{\hbar}\int_{-t/2}^{t/2}dt'\sin{(2\pi 50t' + \beta)}\\ &= \frac{\mu_BgB_0t}{\hbar}\sin{\beta}|\mathrm{sinc}(\pi 50t)|.
\end{align}
We define the contrast $C$ as~\cite{PhysRevLett.110.110503} 
\begin{equation}\label{eq:cont}
    C(t) =\frac{1}{2\pi} \int_0^{2\pi}d\beta\cos(\Phi(\beta))=J_0(|\frac{\mu_BgB_0t}{\hbar}\mathrm{sinc}(\pi 50t)|),
\end{equation}
where $J_0$ is the zeroth Bessel function of the first kind. 

Next, we describe the action of $B(t)$ on the ground-state qubit in terms of a single-qubit dephasing channel with a time-dependent contrast $C(t)$. 
The single-qubit dephasing channel has the form 
\begin{equation}\label{eq:memimp}
    \mathcal{E}(\rho) =\frac{1+C(t)}{2}\rho + \frac{1-C(t)}{2}\sigma_z\rho\sigma_z^\dag.
\end{equation}
The corresponding process matrix in the Pauli basis is given by
\begin{equation*}
    \chi(t) =
	\begin{pmatrix}
		\frac{1+C(t)}{2}&0&0&0\\
		0 &0& 0& 0\\ 
		0 &0&0& 0\\ 
		0&0&0&\frac{1-C(t)}{2}\\
	\end{pmatrix},
\end{equation*}
where the row and column indices $0,1,2,3$ correspond to the Pauli operators $\mathcal{I}$, $\sigma_x$, $\sigma_y$, and $\sigma_z$, respectively. 
The fidelity between $\chi(t)$ and the ideal identity process, $F(t)$ as defined in Section~\ref{sec:ionresults}, is given by $F(t) = (1+C(t))/2$. 

We account for imperfections in state preparation for memory process characterisation by a separate depolarizing channel which, when concatenated with the previous channel, yields the process fidelity model. 
The memory-qubit process fidelity for storage time $t$ is related to that contrast by $F(t) = A'(1 + C(t))/2$, where we have introduced a factor $A'$ to capture imperfections in state preparation. Using Equation~\ref{eq:cont}, the final model for the process fidelity can be expressed as  
\begin{equation}\label{app:eq:proc50}
F(t) = A'(1 + J_0(|\frac{\mu_BgB_0t}{\hbar}\mathrm{sinc}(\pi 50t)|))/2.
\end{equation}

The model has two parameters, $A'$ and $B_0$. How we choose the parameter values used to produce the simulation in the inset of Figure~\ref{fig:ionmem}a is now described. 
The value of $A'$ is chosen such that the process fidelity predicted by Equation~\ref{app:eq:proc50} at time $t = 0$ matches that of Equation~\ref{eq:fid} at $t = 0$. That is achieved for $A' = 0.25 + 0.75A$. The value $A=0.714(5)$ is used, that was obtained from a fit of Equation~\ref{eq:fid} to the data in the main panel of Figure~\ref{fig:ionmem}a., yielding $A' = 0.96$.  

The magnetic field amplitude at \SI{50}{\hertz} ($B_0$) is independently measured to be \SI{36(2)}{\micro\gauss}.
That measurement uses Ramsey experiments on the $\Ss\leftrightarrow\Dd$ transition of a single ion, with an interrogation time of \SI{500}{\micro\second}. 
The Ramsey experiments are synchronized with the \SI{50}{\hertz} magnetic field. 
A wait time before the first $\pi/2$ pulse of the Ramsey experiment is scanned between $0$ and \SI{20}{\milli\second}, changing the phase of the \SI{50}{\hertz} magnetic field incident on the ion during interrogation time. 
An estimate of $B_0$ is obtained from the amplitude of a sinusoidal fit to the data.  
In simulations, the value $B_0 = $~\SI{36}{\micro\gauss} is used. %

\subsection{Depolarization model for a memory protected by dynamical decoupling}\label{appendix:imperfectmemroy}

Equation~\ref{eq:fid} presents a model for the dynamics of the process fidelity of a ground state qubit memory when undergoing dynamical decoupling via the XY16 sequence. 
That model is used to produce the simulated fidelities presented in Figure~\ref{fig:ionmem} (gray lines) and Figure~\ref{fig:robustness} (gray line), often referred to as the `bare memory'. 
The model is based on single-qubit depolarization, which incorporates $\sigma_x-$, $\sigma_y-$ and $\sigma_z-$type errors. Each of those errors could occur due to imperfect state preparation, or due to the combined effect of fluctuations in the ion-qubit frequency and in the power of the electromagnetic field generated by the RF coil at the ions, occurring throughout the dynamical decoupling sequence.
A brief description is now given of how that model is derived. 

A single-qubit depolarization channel~\cite{PhysRevA.70.010304} on a density matrix $\rho$ for time $t$ has the form
\begin{equation}\label{eq:memimp}
    \mathcal{E}(\rho) = (1-p)\rho + p\mathcal{I}/2,
\end{equation}
where $p(t) = 1-e^{-t/\tau}$, $\tau$ is the memory-qubit coherence time, and $\mathcal{I}$ is identity process matrix as defined in Section.~\ref{sec:ionresults}.
The process matrix in the Pauli basis corresponding to that channel is given by
\begin{equation*}
    \chi(t) =
	\begin{pmatrix}
		1-3p/4&0&0&0\\
		0 &p/4& 0& 0\\ 
		0 &0&p/4& 0\\ 
		0&0&0&p/4\\
	\end{pmatrix}.
\end{equation*}
The fidelity between that process matrix and the identity is $F(t) = 0.75e^{-t/\tau} + 0.25$. 

We account for imperfections in state preparation for memory process characterisation by a separate depolarizing channel which, when concatenated with the previous channel, yields the process fidelity model of Equation (1) in the main text. 
\begin{equation}\label{depolfid}
    F(t) = 0.75Ae^{-t/\tau} + 0.25,
\end{equation}
where $0\leq A\leq 1$ is a real number capturing state preparation errors and $0.75A+0.25$ is the initial process fidelity at $t=0$.

\subsection{Depolarization model of ion-photon entanglement infidelity}\label{appendix:imperfectmemroyionphoton}

Section~\ref{ionphotonstoragesec} reported on the storage of ion-photon entanglement in a single ion-memory. The measured state fidelities are presented in Figure~\ref{fig:ionphoton} alongside predictions from a model. The model considers two sources of imperfection: state preparation errors and memory decoherence, as now described.

State preparation errors are modelled by considering the action of separate uncorrelated depolarization channels, each acting on a different qubit (ion and photon) with the same probability  $p_{\mathrm{S}}$, on a perfect Bell state $\ket{\phi}$. The generated imperfectly prepared state $\rho'$ is then given by  
\begin{equation}
\label{humpf}
	\rho' = (1 - p_{\mathrm{S}})^2\ket{\phi}\bra{\phi} + (1 - (1-p_{\mathrm{S}})^2)\mathcal{I}/4.
\end{equation}
The depolarizing channel describing the ion-memory decoherence after a storage time $t$ (Equation~\ref{eq:memimp}) is then applied to the output state Eq. \ref{humpf}, resulting in a final state described analytically by 
\begin{align}
	\rho''(t) &= (1 - p(t))(1 - p_{\mathrm{S}})^2\ket{\phi}\bra{\phi} \\
	&+ (1 - (1 - p(t))(1 - p_{\mathrm{S}})^2)\mathcal{I}/4,
\end{align}
with $p(t) = 1 - e^{-t/\tau}$ describing the probability of a memory error on the ion-qubit. 
Finally, the state fidelity of $\rho''(t)$ with the initial Bell state is given analytically by 
\begin{equation}
\label{arf}
	F_B(t) =  F(\rho''(t), \ket{\phi}) = 0.75Ae^{-t/\tau} + 0.25,
\end{equation}
where $A = (1 - p_{\mathrm{S}})^2$. 
Eq.~\ref{arf} is used to produce the simulated fidelity in Figure~\ref{fig:ionphoton}a, via a fit of the parameters $\tau$ and $A$ to the data.

\subsection{Master equation model of Raman laser crosstalk}\label{appendix:fullmodel}

Equation~\ref{crosstalk} provides a model for the dynamics of the process fidelities of the co-trapped memory-qubits in the presence of photon generation attempts on ion 1. 
That equation is used to produce the simulated process fidelities shown in Figure~\ref{fig:robustness}a (Color-matched shaded regions).
Equation~\ref{crosstalk} models the following imperfections: decoherence of the memory-qubits in the absence of photon generation (`Bare memory', Section~\ref{appendix:imperfectmemroy}); imperfect memory state preparation and Raman laser cross-talk, quantified by the Rabi frequency $\Omega_i$ on ion $i$. 
How Equation~\ref{crosstalk} was obtained is now described.

In summary, we first develop a master equation model for memory-qubit process fidelity. Then we solve it for a range of storage times and levels of crosstalk ($\Omega_i$) and see that the predictions are well described by a simple analytic function. Finally, we modify the function to include imperfect memory state preparation. These steps are now described in more detail. \\

We use the 18-level master equation model, with the ion-cavity coupling strength set to zero ($g=0$). We extend the model to include the master-equation equivalent of the ground-state qubit depolarization channel described in Section~\ref{appendix:imperfectmemroy}---which captures the spin-echo-protected memory decoherence in the absence of photon generation, parameterised by the `bare memory' coherence time $\tau$.  
The initial state of the $\Ca$ ion in the model is prepared in a chosen state $\ket{\psi^{in}_i}$. 
The master equation is solved to calculate the ion state for an evolution time $t$ under illumination by a Raman laser with Rabi frequency $\Omega_i$, yielding the 18-level output state $\rho(t)_{out}$. 
The output state contains population terms in the $\Dd$ and $\Ddt$ manifolds due to off-resonant scattering. 
In the associated experiment, that population is repumped back to the $\Ss$ manifolds before state readout. 
This process is modelled by first taking a partial trace of  $\rho(t)_{out}$ over all levels outside the $\Ss$ manifold, resulting in a normalised qubit state $\rho(t)_{ss'}$.
The final time evolved memory-qubit state is calculated as $\rho(t)_{ss'}=(1-\beta)\rho_{ss'}+\beta\mathcal{I}$, where $\mathcal{I}$ is the identity operator and $\beta$ is the probability that the electron was found in those D-state manifolds. Physically, the second term represents the repumped population being equally distributed between the S-manifold, which is expected to be a reasonable approximation.  Output states $\rho(t)_{ss'}$ are generated for different input states allowing the process matrix $\chi(t)$ to be reconstructed. The fidelity of that matrix is calculated with the identity process. 

Solving the master equation for many seconds of evolution is too intensive given our available computational resources. Therefore, simulations are run for up to \SI{5}{\milli\second} and repeated for a suitable range of Rabi frequencies $\Omega_i$ and `bare memory' coherence times $\tau$. In all cases, predicted fidelities are seen to decay exponentially with storage time $t$, and are well described by the anstatz function 
\begin{equation}
\label{f250}
F_i(t) = e^{-t/\tau}e^{0.67\gamma^{\mathrm{eff}}_i t}+0.25, 
\end{equation}
where $\gamma^{\mathrm{eff}}_i = \gamma(\Omega_i/2\Delta)^2$ is the effective scattering rate, $\gamma = 23$MHz~\cite{PhysRevLett.70.3213} is the spontaneous scattering rate of the $\Pp$ manifold, and $\Delta = $~\SI{400}{\mega\hertz} is the laser detuning from the $\Ss\leftrightarrow\Pp$ transition.

Two final parameters are required to model the experiment that produced the results shown in Figure~\ref{fig:robustness}a. First, a factor $B$ is introduced to account for the fraction of the storage time during which the Raman laser is applied within the experimental sequence, which here is $B=0.35$. Second, imperfections in the preparation of the ground-state qubit are incorporated, as in Section~\ref{appendix:imperfectmemroy}, through a concatenated depolarization channel parameterised by a factor $A$, with ($0\leq A \leq 1$). This leads to Equation~\ref{crosstalk}, reproduced here for convenience:
\begin{equation}\label{app:decayrate}
	F_i(t) = Ae^{-t/\tau}e^{0.67\gamma^{\mathrm{eff}}_iB t}+0.25.
\end{equation}

\section{Robustness of XY16 sequence to 50 HZ noise}\label{appendix:50hzxy16}

Here we present a model of how measured magnetic field noise at \SI{50}{\hertz} affects the coherence of ground state memory qubits during the XY16 dynamical coupling sequence. The model predicts no significant effect for multisecond storage times. 

The XY16 spin echo sequence ideally consists of 16 $\pi$ pulses, each of duration $t_{\mathrm{pulse}}$ and equally spaced by wait time of $t_\mathrm{w}$, with phases following the list $\Phi_{XY16}$ in Equation~\ref{eq:XY16phases}. 
That sequence is repeated to achieve the desired storage time, involving a total of $N_{\mathrm{tot}}$ spin echo pulses. 
The \SI{50}{\hertz} magnetic field is described by Equation~\ref{50hzb}, with an amplitude $B_0$ and a randomly-fluctuating phase $\beta$. 
In a single run of the model (shot), the phase is considered to be fixed. 
The magnetic field is modelled to interact with the ground-state qubit via the linear Zeeman Hamiltonian defined in Equation~\ref{linzee}. 
 That interaction causes a time-varying frequency detuning between the applied spin echo radio frequency (RF) and the $\ket{S}\leftrightarrow\ket{S'}$ transition of the form 
\begin{equation}
	\Delta(t') = \Delta_0 + \frac{\mu_BgB_0}{2\hbar}\sin{(2\pi50t' + \beta)},
\end{equation}
where $g = 2$ is the Land\'e factor for the $\Ss$ manifold, $\mu_B$ is the Bohr magneton, $\Delta_0$ is a fixed frequency mismatch between the RF drive and the resonance frequency of the $\ket{S}\leftrightarrow\ket{S'}$ transition due to a calibration error, and $t'$ is time within the single shot.

The detuning  $\Delta(t')$ results in an imperfection in the $n$-th spin echo $\pi$ pulse. The imperfect action of the $n^{\mathrm{th}}$ spin echo is modelled via the unitary rotation~\cite{Foot2005, Cohen_Tannoudji_atomphoton}
\begin{equation}
R_n(t_\mathrm{pulse}, \phi_{n|16}) = e^{-i\frac{\pi t_{\mathrm{pulse}}}{2t_{\pi}}\bm{\sigma}\cdot\bm{c}},
\end{equation}
where we have introduced the Pauli vector $\bm{\sigma} = (\sigma_x, \sigma_y, \sigma_z)$, the rotation axis $\bm{c}$, and the ideal spin-echo $\pi$ time $t_{\pi}$. 
The rotation axis can be expressed as $\bm{c_n} = (\Omega\cos{\phi_{n|16}}, \Omega\sin{\phi_{n|16}}, \Delta(t_n))/\Omega_{\mathrm{eff}}$, where $\phi_{n|16}$ is the  $\mathrm{mod}_{16}(n)$-th element of the list of phases $\Phi_{XY16}$, $\Omega_{\mathrm{eff}} = \sqrt{\Omega^2 +\Delta(t_n)^2}$ is the effective Rabi frequency and $\Omega$ is the Rabi frequency of the RF drive.  
Here we have introduced $\Delta(t_n)$, which is the detuning at the time of the middle of the $n^{\mathrm{th}}$ spin echo. 
The $\pi$ time can be expressed as $t_{\pi} = \frac{\pi}{\Omega_{\mathrm{eff}}}$.%

During the wait time that occurs between the $n^{\mathrm{th}}$ and $(n+1)^{\mathrm{th}}$ spin echo, the detuning causes a $\sigma_z$ rotation of the form
\begin{equation}
W_n(t_{\mathrm{w}}) = e^{-i\Delta(\frac{t_{n+1}-t_{n}}{2}) t_{\mathrm{w}}\sigma_z/2},
\end{equation}
where the relevant detuning $\Delta$ is fixed to a constant value, defined by its value at the middle of the wait time. %

The effect of the \SI{50}{\hertz} magnetic field fluctuation on the whole sequence, where imperfections in each spin echo and wait time accumulate, are modelled by the following sequence of unitary rotations
\begin{equation} \label{keymetho:eq:unitaryxy16}
U_{XY16} = W_n(t_{\mathrm{w}})\prod_n^{N_{\mathrm{tot}}} R_n(t_\mathrm{pulse}, \phi_{n|16})W_n(t_{\mathrm{w}}).
\end{equation}
Fifty unitaries $U_{XY16}$ are found, each using different a phase value $\beta$ of the \SI{50}{\hertz} magnetic field fluctuation sampled from a random uniform distribution. 
From each of the fifty unitaries, the corresponding process matrix is determined, and the process fidelity of that process matrix with $\mathcal{I}$ is calculated. 
Those fifty process fidelities are then averaged. 

We use the above model to calculate the process fidelity for $N_{\mathrm{tot}} = 10240$, corresponding to a two second storage time, and for the measured value $B_0 = $~\SI{36}{\micro\gauss}, as described in Section~\ref{appendix:dephase}. 
The calculation is repeated for miscalibrations in spin echo RF detuning over the range $\Delta_0 \in [-500, 500]$~\SI{}{\hertz}, with $t_{\mathrm{pulse}}$ set to the ideal $\pi$ time in the absence of miscalibration and magnetic field noise ($t_{\pi}$ for $\Delta(t')=0$) . 
The calculation is repeated again for miscalibrations in spin echo pulse durations over the range $t_{\mathrm{pulse}}/t_{\pi} \in [-5, 5]$~\%. 
Both ranges are at least a factor of ten larger than we expect them to be in the experiment, as discussed in Section~\ref{appendix:xy16robustness}. 
Over all cases considered (calculations done) the process fidelity reduces by at most $0.006$. We therefore conclude that  \SI{50}{\hertz} magnetic field noise, calibration errors in spin echo pulse lengths and frequencies are expected to have no significant effect on the coherence of ground state memory qubits during XY16 dynamical coupling.\\

\section{Calibration of spin-echo pulses and Robustness of the XY16 sequence to miscalibrations}\label{appendix:xy16robustness}

This section first describes our procedure for calibrating and setting the pulse length and frequency used for the RF spin echoes in the XY16 sequence. Second, our calibration uncertainties are compared to a simulation predicting their effect on the process fidelity of the XY16 sequence, leading to the conclusion that they have no significant effect on the ground-state memory process fidelity over five seconds.

The experimental sequence for spin-echo calibration begins with Doppler cooling, followed by optical pumping of the ions into the state $\ket{S}$. An RF pulse then ideally prepares each ion in the state $\frac{\ket{S} + \ket{S'}}{\sqrt{2}}$.
We then repeatedly apply the XY4 sequence~\cite{deLange:2010avk}, implemented with RF pulses. XY4---a variant of XY16 that is less robust to miscalibration of $t_{\mathrm{pulse}}$ and the spin-echo frequency $f$---consists of four spin-echo pulses of duration $t_{\mathrm{pulse}}$ separated by equally spaced wait times $t_\mathrm{w}$. In total, the repeated XY4 sequence uses $N_{\mathrm{tot}} = 480$ pulses, corresponding to a storage time of~\SI{100}{\milli\second}.
Finally, each ion is measured in the $\sigma_x$ basis by applying an RF $\pi/2$ pulse, followed by an optical $\pi$ pulse that maps the $\ket{S'}$ state into the $\Dd$ manifold, and then performing fluorescence detection to obtain ion excitation probabilities.

First, we repeat the experimental sequence while sweeping the RF frequency $f$ over $\pm\SI{1}{\kilo\hertz}$ around the nominal (current best estimate) resonance frequency, with $t_{\mathrm{pulse}}$ fixed at the nominal $\pi$ time.
Then, we repeat the sequence while sweeping the pulse length $t_{\mathrm{pulse}}$ over $\pm\SI{15}{\micro\second}$ around the nominal $\pi$ time, with the RF frequency fixed at the nominal resonance frequency.

We now describe how estimates of miscalibrations in the nominal pulse duration and frequency of the spin echoes are extracted from the excitation probabilities.
We employ the model in Equation~\ref{keymetho:eq:unitaryxy16} for the ideal unitary evolution of the XY16 sequence, with the modification that the phases follow the XY4 sequence, i.e., the first four elements of the list $\Phi_{XY16}$ in Equation~\ref{eq:XY16phases}.
Magnetic-field fluctuations at \SI{50}{\hertz} are neglected by setting $B_0 = 0$.
The frequency miscalibration is quantified by $\Delta_0 = f - f_R$, where $f_R$ is the resonance frequency of the $\ket{S}\leftrightarrow\ket{S'}$ transition.
The pulse-duration miscalibration is quantified by $t_{\mathrm{pulse}}/t_{\pi}$, where $t_{\pi}$ is the correct $\pi$ time.
From the unitary $U_{XY4}$ generated by the model, we predict the excitation probability via
$P_{\mathrm{exc}} = \left|\bra{1} U_{XY4} \ket{0}\right|^2$.
The parameters $\Delta_0$ and $t_{\mathrm{pulse}}$ are chosen by eye to provide the closest fit to the measured excitation probabilities. The estimated uncertainty in $\Delta_0$ is $\pm$\SI{20}{\hertz} and $\pm 0.002$ in $t_{\mathrm{pulse}}/t_{\pi}$.

\begin{figure}[t!]
	\begin{center}
       \includegraphics[width=1\columnwidth]{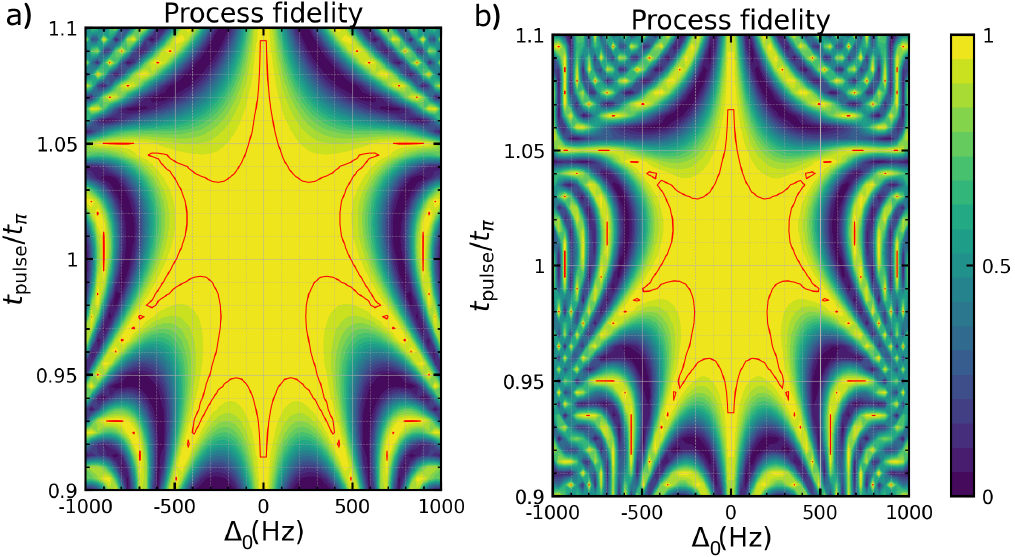}
		\caption{Simulated performance of XY16 sequence in the presence of RF spin echo miscalibrations. (a),(b) Process fidelity between simulated XY16 process and ideal identity process for storage times of \SI{2}{\second} and \SI{5}{\second}, respectively, with parameters described in the text. 
		Frequency miscalibrations are a detuning ($\Delta_0 = f - f_R$) of the RF frequency ($f$) from the correct resonance frequency ($f_R$) of the $\ket{S}\leftrightarrow\ket{S'}$ transition. 
		Fractional pulse duration miscalibrations are $t_{\mathrm{pulse}}/t_{\pi}$, where $t_{\mathrm{pulse}}$ is the duration of each spin echo pulse, and $t_{\pi}$ is the correct $\pi$-time. 
		Red lines are the 0.99 contour.}\label{fig:approbustness}
		\vspace{-5 mm}
	\end{center}
\end{figure}

The robustness of the XY16 sequence to spin-echo miscalibrations is investigated by repeating the above simulation with phases following the XY16 sequence (Equation~\ref{eq:XY16phases}) and for storage times of \SI{2}{\second} and \SI{5}{\second}.
For the \SI{2}{\second} storage time, which corresponds to $N_{\mathrm{tot}} = 10240$ total spin echoes, the simulated process fidelity exceeds $0.99$ over a square region defined by $\Delta_0 \in \pm\SI{220}{\hertz}$ and $t_{\mathrm{pulse}}/t_{\pi} \in 1 \pm 0.035$ (Figure~\ref{fig:approbustness}a, yellow central region).
For the \SI{5}{\second} storage time, which corresponds to $N_{\mathrm{tot}} = 25600$ total spin echoes, that region is reduced to $\Delta_0 \in \pm\SI{200}{\hertz}$ and $t_{\mathrm{pulse}}/t_{\pi} \in 1 \pm 0.03$ (Figure~\ref{fig:approbustness}b, smaller yellow central region).
Since our estimated miscalibrations from the XY4 procedure lie well within these regions for both storage times, we conclude that they affect the process fidelity of the XY16 sequence only at the sub-percent level for storage times up to \SI{5}{\second}.

\section{Composite X5 sequence error}\label{appendix:imperfectpi}

A composite $\pi$ pulse---an X5 sequence---was used in experiments investigating memory robustness in the five-ion chain (Section~\ref{robustnesssec}). 
The X5 sequence is described in step 15 of the list in Appendix~\ref{appendix:pulsesequence}, and ideally maps $\ket{S'}$ to $\ket{D'}$, making $\ket{S}$ and $\ket{S'}$ fully distinguishable via fluorescence detection. 
Errors in the X5 sequence can introduce errors into experimentally reconstructed process matrices, if they are not taken into account. 
Here, we first describe the experiments done to quantify the size of errors in our X5 sequences and give the results, which depend on the ion-string temperature. 
Second, we show how the process matrix reconstruction can be modified to take these errors into account. 
Third, we find that when taking into account X5 sequence errors in this way, process fidelities are obtained from the data that are not statistically different when X5 sequence errors are not considered. As such, X5 sequence errors have no statistically significant effect on the process fidelities. None of the process fidelities presented in the main text take X5 sequence errors into account.

The experimental sequence used to measure the error of the composite X5 sequence is performed on a five-ion string with the same trapping parameters as used to obtain the results presented in Section~\ref{robustnesssec}. After Doppler cooling and optically pumping each ion into the state $\ket{S}$, an RF $\pi$ pulse is applied to transfer each ion to $\ket{S'}$.
Errors in that RF $\pi$ pulse are expected to be negligible and are not considered.
The five \SI{729}{\nano\meter} $\pi$ pulses of the composite X5 sequence (with phases defined in Equation~\ref{eq:X5phases}) are then applied to the $\ket{S'}\leftrightarrow\ket{D'}$ transition using the \SI{729}{\nano\meter} laser.
Fluorescence detection of each ion is performed with the camera, completing a single shot of the experimental sequence.
The experimental sequence is repeated for $N$ shots.
The error in the composite X5 sequence is quantified by
$$
p_{\mathrm{err}} = \frac{1}{5N}\sum_{i=1}^{5} N_i^{\mathrm{bright}},
$$
where $N_i^{\mathrm{bright}}$ is the total number of times that the $i$-th ion is observed to be bright over all $N$ shots. Thus, $p_{\mathrm{err}}$ is the average X5-sequence error over the five-ion chain.
The statistical uncertainty in $p_{\mathrm{err}}$ due to shot noise is estimated as
$$
\sqrt{\frac{p_{\mathrm{err}} \left(1 - p_{\mathrm{err}}\right)}{5N}}.
$$
The experimental result is $p_{\mathrm{err}} = 0.041(6)$. This value will be used for the X5 sequence error for a Doppler cooled ion string, in subsequent calculations. 

The same procedure is repeated after making two changes. First,  after preparing all the ions in the $\ket{S}$ we now include 8640 monochromatic photon generation attempts on ion 1. The repeated photon generation attempts heat the motional modes of the ion string via photon recoil, which in turn modifies the coupling of the \SI{729}{\nano\meter} laser to the ion transition.  Second, after those attempts, the ions are optically pumped back into $\ket{S}$, before the single RF $\pi$ pulse is applied which transfers the electronic state of each ion to $\ket{S'}$, as described above. The experimental result in this case is $p_{\mathrm{err}} = 0.093(8)$.  

Next, the process matrix reconstruction is modified to take the X5 sequence error into account. 
When using an imperfect $\pi$ pulse to map the bright $\ket{S'}$ state of the ground state qubit to the dark $\ket{D'}$ state, electronic population is erroneously left in the bright $\ket{S'}$ state with probability $p_{\mathrm{err}}$. 
Specifically, the state transformation realised is $\ket{S'}\bra{S'}\rightarrow p_{\mathrm{err}}\ket{S'}\bra{S'} + (1-p_{\mathrm{err}})\ket{D'}\bra{D'}$. 
When electron population is erroneously left in the bright $\ket{S'}$, fluorescence detection does not distinguish it from the bright $\ket{S}$ state. 
We correct for that error by adjusting the projectors used in the process reconstruction associated with each measurement outcome. Each projector is in a basis corresponding to a different Pauli operator $\sigma_\alpha$, with $\alpha \in [x, y, z]$. 
The unmodified projectors correspond to the eigenvectors of the corresponding Pauli operator: $\ket{\downarrow_\alpha}\bra{\downarrow_\alpha}$ and $\ket{\uparrow_\alpha}\bra{\uparrow_\alpha}$. 
Modified projectors, corrected for an error of size $p_{\mathrm{err}}$, are given by $\ket{\downarrow_\alpha}\bra{\downarrow_\alpha} + p_{\mathrm{err}}\ket{\uparrow_\alpha}\bra{\uparrow_\alpha}$ and $(1-p_{\mathrm{err}})\ket{\uparrow_\alpha}\bra{\uparrow_\alpha}$.

Recall that Figure~\ref{fig:robustness}a shows processes fidelities reconstructed for four memory ions in the presence of photon generation attempts (from 160 to 8640) without considering X5 sequence error. For 160 attempts we expect that the ion string temperature is not significantly different from that after intial Doppler cooling and thus expect an X5 sequence error of $p_{\mathrm{err}} = 0.041(6)$. When using the corresponding modified process reconstruction to obtain process fidelities for the data taken after 160 attempts, the values are larger than the corresponding ones in Figure~\ref{fig:robustness}a by an amount that is smaller than one standard deviation of uncertainty. For 8640 attempts we expect an X5 sequence error of $p_{\mathrm{err}} = 0.093(8)$. When using the corresponding modified process reconstruction to obtain process fidelities for the data taken after 8640 attempts, the values are larger than the corresponding ones in Figure~\ref{fig:robustness}a by an amount that is again smaller than one standard deviation of uncertainty. We therefore conclude that the effect of X5 sequence error is minimal and do not consider it in any of the results presented in the main text.

\bibliography{bibliography}

@article{hvsx-cx2d,
	author = {Cussenot, P. and Grivet, B. and Feldmann, L. and Wengerowsky, S. and Lanyon, B. P. and Northup, T. E. and de Riedmatten, H. and S\o{}rensen, A. S. and Sangouard, N.},
	doi = {10.1103/hvsx-cx2d},
	issue = {24},
	journal = {Phys. Rev. Lett.},
	month = {Dec},
	numpages = {9},
	pages = {240803},
	publisher = {American Physical Society},
	title = {Uniting Quantum Processing Nodes of Cavity-Coupled Ions with Rare-Earth Quantum Repeaters Using Single-Photon Pulse Shaping Based on Atomic Frequency Comb},
	url = {https://link.aps.org/doi/10.1103/hvsx-cx2d},
	volume = {135},
	year = {2025}}

@article{PhysRevA.62.052317,
	author = {Eisert, J. and Jacobs, K. and Papadopoulos, P. and Plenio, M. B.},
	doi = {10.1103/PhysRevA.62.052317},
	issue = {5},
	journal = {Phys. Rev. A},
	month = {Oct},
	numpages = {7},
	pages = {052317},
	publisher = {American Physical Society},
	title = {Optimal local implementation of nonlocal quantum gates},
	url = {https://link.aps.org/doi/10.1103/PhysRevA.62.052317},
	volume = {62},
	year = {2000}}

@article{Gottesman:1999oe,
	author = {Gottesman, Daniel and Chuang, Isaac L.},
	date = {1999/11/01},
	doi = {10.1038/46503},
	id = {Gottesman1999},
	isbn = {1476-4687},
	journal = {Nature},
	number = {6760},
	pages = {390--393},
	title = {Demonstrating the viability of universal quantum computation using teleportation and single-qubit operations},
	url = {https://doi.org/10.1038/46503},
	volume = {402},
	year = {1999}}

@article{PhysRevLett.88.217901,
	author = {Ekert, Artur K. and Alves, Carolina Moura and Oi, Daniel K. L. and Horodecki, Micha\l{} and Horodecki, Pawe\l{} and Kwek, L. C.},
	doi = {10.1103/PhysRevLett.88.217901},
	issue = {21},
	journal = {Phys. Rev. Lett.},
	month = {May},
	numpages = {4},
	pages = {217901},
	publisher = {American Physical Society},
	title = {Direct Estimations of Linear and Nonlinear Functionals of a Quantum State},
	url = {https://link.aps.org/doi/10.1103/PhysRevLett.88.217901},
	volume = {88},
	year = {2002}}

@article{PhysRevLett.81.5932,
	author = {Briegel, H.-J. and D\"ur, W. and Cirac, J. I. and Zoller, P.},
	doi = {10.1103/PhysRevLett.81.5932},
	issue = {26},
	journal = {Phys. Rev. Lett.},
	month = {Dec},
	numpages = {0},
	pages = {5932--5935},
	publisher = {American Physical Society},
	title = {Quantum Repeaters: The Role of Imperfect Local Operations in Quantum Communication},
	url = {https://link.aps.org/doi/10.1103/PhysRevLett.81.5932},
	volume = {81},
	year = {1998}}

@online{PreskillPH229-Ch3,
	author = {Preskill, John},
	organization = {California Institute of Technology},
	title = {Chapter 3: Quantum Teleportation (Lecture Notes for Physics 229)},
	url = {http://theory.caltech.edu/~preskill/ph229/ph229_ch3_teleportation.pdf},
	urldate = {2026-04-26},
	year = {1998}}

@article{PhysRevLett.70.1895,
	author = {Bennett, Charles H. and Brassard, Gilles and Cr\'epeau, Claude and Jozsa, Richard and Peres, Asher and Wootters, William K.},
	doi = {10.1103/PhysRevLett.70.1895},
	issue = {13},
	journal = {Phys. Rev. Lett.},
	month = {Mar},
	numpages = {0},
	pages = {1895--1899},
	publisher = {American Physical Society},
	title = {Teleporting an unknown quantum state via dual classical and Einstein-Podolsky-Rosen channels},
	url = {https://link.aps.org/doi/10.1103/PhysRevLett.70.1895},
	volume = {70},
	year = {1993}}

@article{Ong_2020,
	author = {Ong, Florian R and Sch{\"u}ppert, Klemens and Jobez, Pierre and Teller, Markus and Ames, Ben and Fioretto, Dario A and Friebe, Konstantin and Lee, Moonjoo and Colombe, Yves and Blatt, Rainer and Northup, Tracy E},
	doi = {10.1088/1367-2630/ab8af9},
	journal = {New Journal of Physics},
	month = {jun},
	number = {6},
	pages = {063018},
	publisher = {IOP Publishing},
	title = {Probing surface charge densities on optical fibers with a trapped ion},
	url = {https://doi.org/10.1088/1367-2630/ab8af9},
	volume = {22},
	year = {2020}}

@article{PhysRevLett.118.250502,
	author = {Inlek, I. V. and Crocker, C. and Lichtman, M. and Sosnova, K. and Monroe, C.},
	doi = {10.1103/PhysRevLett.118.250502},
	issue = {25},
	journal = {Phys. Rev. Lett.},
	month = {Jun},
	numpages = {5},
	pages = {250502},
	publisher = {American Physical Society},
	title = {Multispecies Trapped-Ion Node for Quantum Networking},
	url = {https://link.aps.org/doi/10.1103/PhysRevLett.118.250502},
	volume = {118},
	year = {2017}}

@article{Monroe_2013,
	author = {Monroe, C. and Kim, J.},
	doi = {10.1126/science.1231298},
	issn = {1095-9203},
	journal = {Science},
	month = mar,
	number = {6124},
	pages = {1164--1169},
	publisher = {American Association for the Advancement of Science (AAAS)},
	title = {Scaling the Ion Trap Quantum Processor},
	url = {http://dx.doi.org/10.1126/science.1231298},
	volume = {339},
	year = {2013}}

@article{10.1063/5.0069544,
	author = {Allcock, D. T. C. and Campbell, W. C. and Chiaverini, J. and Chuang, I. L. and Hudson, E. R. and Moore, I. D. and Ransford, A. and Roman, C. and Sage, J. M. and Wineland, D. J.},
	doi = {10.1063/5.0069544},
	issn = {0003-6951},
	journal = {Applied Physics Letters},
	month = {11},
	number = {21},
	pages = {214002},
	title = {omg blueprint for trapped ion quantum computing with metastable states},
	url = {https://doi.org/10.1063/5.0069544},
	volume = {119},
	year = {2021}}

@article{Cui_2025,
	author = {Cui, Z.-B. and Wang, Z.-Q. and Lai, P.-C. and Wang, Y. and Shi, J.-X. and Liu, P.-Y. and Sun, Y.-D. and Tian, Z.-C. and Liang, Y.-B. and Qi, B.-X. and Huang, Y.-Y. and Zhou, Z.-C. and Wu, Y.-K. and Xu, Y. and Duan, L.-M. and Pu, Y.-F.},
	doi = {10.1038/s41467-025-67311-5},
	issn = {2041-1723},
	journal = {Nature Communications},
	month = dec,
	number = {1},
	publisher = {Springer Science and Business Media LLC},
	title = {Metropolitan-scale ion-photon entanglement via a quantum network node with hybrid multiplexing enhancements},
	url = {http://dx.doi.org/10.1038/s41467-025-67311-5},
	volume = {17},
	year = {2025}}

@article{PhysRevLett.134.070801,
	author = {Lai, P.-C. and Wang, Y. and Shi, J.-X. and Cui, Z.-B. and Wang, Z.-Q. and Zhang, S. and Liu, P.-Y. and Tian, Z.-C. and Sun, Y.-D. and Chang, X.-Y. and Qi, B.-X. and Huang, Y.-Y. and Zhou, Z.-C. and Wu, Y.-K. and Xu, Y. and Pu, Y.-F. and Duan, L.-M.},
	doi = {10.1103/PhysRevLett.134.070801},
	issue = {7},
	journal = {Phys. Rev. Lett.},
	month = {Feb},
	numpages = {8},
	pages = {070801},
	publisher = {American Physical Society},
	title = {Realization of a Crosstalk-Free Two-Ion Node for Long-Distance Quantum Networking},
	url = {https://link.aps.org/doi/10.1103/PhysRevLett.134.070801},
	volume = {134},
	year = {2025}}

@article{Feng:2024fy,
	author = {Feng, L. and Huang, Y. -Y. and Wu, Y. -K. and Guo, W. -X. and Ma, J. -Y. and Yang, H. -X. and Zhang, L. and Wang, Y. and Huang, C. -X. and Zhang, C. and Yao, L. and Qi, B. -X. and Pu, Y. -F. and Zhou, Z. -C. and Duan, L. -M.},
	date = {2024/01/03},
	doi = {10.1038/s41467-023-44220-z},
	id = {Feng2024},
	isbn = {2041-1723},
	journal = {Nature Communications},
	number = {1},
	pages = {204},
	title = {Realization of a crosstalk-avoided quantum network node using dual-type qubits of the same ion species},
	url = {https://doi.org/10.1038/s41467-023-44220-z},
	volume = {15},
	year = {2024}}

@article{Yang:2022il,
	author = {Yang, H. -X. and Ma, J. -Y. and Wu, Y. -K. and Wang, Y. and Cao, M. -M. and Guo, W. -X. and Huang, Y. -Y. and Feng, L. and Zhou, Z. -C. and Duan, L. -M.},
	date = {2022/09/01},
	doi = {10.1038/s41567-022-01661-5},
	id = {Yang2022},
	isbn = {1745-2481},
	journal = {Nature Physics},
	number = {9},
	pages = {1058--1061},
	title = {Realizing coherently convertible dual-type qubits with the same ion species},
	url = {https://doi.org/10.1038/s41567-022-01661-5},
	volume = {18},
	year = {2022}}

@book{nielsen_chuang_2010,
	address = {Cambridge, UK},
	author = {Nielsen, Michael A. and Chuang, Isaac L.},
	edition = {10th Anniversary Edition},
	isbn = {978-1107002173},
	publisher = {Cambridge University Press},
	title = {Quantum Computation and Quantum Information},
	year = {2010}}

@article{Covey:2023qy,
	author = {Covey, Jacob P. and Weinfurter, Harald and Bernien, Hannes},
	date = {2023/09/16},
	doi = {10.1038/s41534-023-00759-9},
	id = {Covey2023},
	isbn = {2056-6387},
	journal = {npj Quantum Information},
	number = {1},
	pages = {90},
	title = {Quantum networks with neutral atom processing nodes},
	url = {https://doi.org/10.1038/s41534-023-00759-9},
	volume = {9},
	year = {2023}}

@INPROCEEDINGS{5438603,
  author={Broadbent, Anne and Fitzsimons, Joseph and Kashefi, Elham},
  booktitle={2009 50th Annual IEEE Symposium on Foundations of Computer Science}, 
  title={Universal Blind Quantum Computation}, 
  year={2009},
  volume={},
  number={},
  pages={517-526},
  doi={10.1109/FOCS.2009.36}}

@misc{seubert2026efficiententanglementremotesingleatom,
      title={Efficient entanglement of three remote single-atom quantum-network nodes}, 
      author={Matthias Seubert and Leonardo Ruscio and Tobias Frank and Philip Thomas and Maya Büki and Gianvito Chiarella and Pau Farrera and Olivier Morin and Gerhard Rempe},
      year={2026},
      eprint={2606.32006},
      archivePrefix={arXiv},
      primaryClass={quant-ph},
      url={https://arxiv.org/abs/2606.32006}, 
}

@article{10.1007/s00224-018-9872-3,
author = {Gheorghiu, Alexandru and Kapourniotis, Theodoros and Kashefi, Elham},
title = {Verification of Quantum Computation: An Overview of Existing Approaches},
year = {2019},
issue_date = {May 2019},
publisher = {Springer-Verlag},
address = {Berlin, Heidelberg},
volume = {63},
number = {4},
issn = {1432-4350},
url = {https://doi.org/10.1007/s00224-018-9872-3},
doi = {10.1007/s00224-018-9872-3},
journal = {Theor. Comp. Sys.},
month = may,
pages = {715–808},
numpages = {94}
}

@article{Kirchmair:2009kh,
	author = {Kirchmair, G. and Z{\"a}hringer, F. and Gerritsma, R. and Kleinmann, M. and G{\"u}hne, O. and Cabello, A. and Blatt, R. and Roos, C. F.},
	date = {2009/07/01},
	doi = {10.1038/nature08172},
	id = {Kirchmair2009},
	isbn = {1476-4687},
	journal = {Nature},
	number = {7254},
	pages = {494--497},
	title = {State-independent experimental test of quantum contextuality},
	url = {https://doi.org/10.1038/nature08172},
	volume = {460},
	year = {2009}}

@article{doi:10.1126/science.aad9480,
	author = {Thomas Monz and Daniel Nigg and Esteban A. Martinez and Matthias F. Brandl and Philipp Schindler and Richard Rines and Shannon X. Wang and Isaac L. Chuang and Rainer Blatt},
	doi = {10.1126/science.aad9480},
	journal = {Science},
	number = {6277},
	pages = {1068-1070},
	title = {Realization of a scalable Shor algorithm},
	volume = {351},
	year = {2016}}

@article{Bugalho2025privaterobuststates,
	author = {Bugalho, Lu{\'{i}}s and Hassani, Majid and Omar, Yasser and Markham, Damian},
	doi = {10.22331/q-2025-01-15-1596},
	issn = {2521-327X},
	journal = {{Quantum}},
	month = jan,
	pages = {1596},
	publisher = {{Verein zur F{\"{o}}rderung des Open Access Publizierens in den Quantenwissenschaften}},
	title = {Private and {R}obust {S}tates for {D}istributed {Q}uantum {S}ensing},
	url = {https://doi.org/10.22331/q-2025-01-15-1596},
	volume = {9},
	year = {2025}}

@article{Ruster2016,
	author = {Ruster, T. and Schmiegelow, C. T. and Kaufmann, H. and Warschburger, C. and Schmidt-Kaler, F. and Poschinger, U. G.},
	day = {23},
	doi = {10.1007/s00340-016-6527-4},
	issn = {1432-0649},
	journal = {Applied Physics B},
	month = {Sep},
	number = {10},
	pages = {254},
	title = {A long-lived Zeeman trapped-ion qubit},
	url = {https://doi.org/10.1007/s00340-016-6527-4},
	volume = {122},
	year = {2016}}

@article{repeater,
	author = {Krutyanskiy, V. and Canteri, M. and Meraner, M. and Bate, J. and Krcmarsky, V. and Schupp, J. and Sangouard, N. and Lanyon, B. P.},
	doi = {10.1103/PhysRevLett.130.213601},
	issue = {21},
	journal = {Phys. Rev. Lett.},
	month = {May},
	numpages = {7},
	pages = {213601},
	publisher = {American Physical Society},
	title = {Telecom-Wavelength Quantum Repeater Node Based on a Trapped-Ion Processor},
	url = {https://link.aps.org/doi/10.1103/PhysRevLett.130.213601},
	volume = {130},
	year = {2023}}

@phdthesis{JosefThesis,
	author = {Josef Schupp},
	langid = {english},
	school = {University of Innsbruck},
	title = {Interface between trapped-ion qubits and travelling photons with close-to-optimal efficiency},
	url = {https://www.quantumoptics.at/images/publications/dissertation/Schupp_thesis.pdf}}

@article{101,
	author = {Krutyanskiy, V. and Canteri, M. and Meraner, M. and Krcmarsky, V. and Lanyon, B.P.},
	doi = {10.1103/PRXQuantum.5.020308},
	issue = {2},
	journal = {PRX Quantum},
	month = {Apr},
	numpages = {11},
	pages = {020308},
	publisher = {American Physical Society},
	title = {Multimode Ion-Photon Entanglement over 101 Kilometers},
	url = {https://link.aps.org/doi/10.1103/PRXQuantum.5.020308},
	volume = {5},
	year = {2024}}

@book{Foot2005,
	author = {Foot, CJ},
	publisher = {Oxford University Press, USA},
	title = {Atomic physics},
	url = {http://scholar.google.com/scholar.bib?q=info:RG8FFqhmF0kJ:scholar.google.com/&output=citation&hl=en&ct=citation&cd=0},
	year = 2005}

@book{Cohen_Tannoudji_atomphoton,
	address = {New York},
	author = {{Cohen-Tannoudji}, C. and {Grynberg}, G. and {Dupont-Roc}, J.},
	publisher = {Wiley},
	title = {Atom-Photon Interactions: Basic Processes and Applications},
	year = 1992}

@article{3hgx-wcdn,
	author = {Bate, J. and Hamann, A. and Canteri, M. and Winkler, A. and Koong, Z. X. and Krutyanskiy, V. and D\"ur, W. and Lanyon, B. P.},
	doi = {10.1103/3hgx-wcdn},
	issue = {22},
	journal = {Phys. Rev. Lett.},
	month = {Nov},
	numpages = {8},
	pages = {220801},
	publisher = {American Physical Society},
	title = {Experimental Distributed Quantum Sensing in a Noisy Environment},
	url = {https://link.aps.org/doi/10.1103/3hgx-wcdn},
	volume = {135},
	year = {2025}}

@article{PhysRevLett.110.110503,
	author = {Kotler, Shlomi and Akerman, Nitzan and Glickman, Yinnon and Ozeri, Roee},
	doi = {10.1103/PhysRevLett.110.110503},
	issue = {11},
	journal = {Phys. Rev. Lett.},
	month = {Mar},
	numpages = {5},
	pages = {110503},
	publisher = {American Physical Society},
	title = {Nonlinear Single-Spin Spectrum Analyzer},
	url = {https://link.aps.org/doi/10.1103/PhysRevLett.110.110503},
	volume = {110},
	year = {2013}}

@article{security,
	author = {Hassani, Majid and Scheiner, Santiago and Paris, Matteo and Markham, Damian},
	doi = {10.1103/PhysRevLett.134.030802},
	journal = {Physical Review Letters},
	month = {01},
	title = {Privacy in Networks of Quantum Sensors},
	volume = {134},
	year = {2025}}

@article{Friis2019,
	author = {Friis, Nicolai and Vitagliano, Giuseppe and Malik, Mehul and Huber, Marcus},
	day = {01},
	doi = {10.1038/s42254-018-0003-5},
	issn = {2522-5820},
	journal = {Nature Reviews Physics},
	month = {Jan},
	number = {1},
	pages = {72-87},
	title = {Entanglement certification from theory to experiment},
	url = {https://doi.org/10.1038/s42254-018-0003-5},
	volume = {1},
	year = {2019}}

@article{PhysRevA.70.010304,
	author = {Daffer, Sonja and W\'odkiewicz, Krzysztof and Cresser, James D. and McIver, John K.},
	doi = {10.1103/PhysRevA.70.010304},
	issue = {1},
	journal = {Phys. Rev. A},
	month = {Jul},
	numpages = {4},
	pages = {010304},
	publisher = {American Physical Society},
	title = {Depolarizing channel as a completely positive map with memory},
	url = {https://link.aps.org/doi/10.1103/PhysRevA.70.010304},
	volume = {70},
	year = {2004}}

@mastersthesis{MarcoThesis,
	author = {Marco Canteri},
	school = {University of Innsbruck},
	title = {Single-atom-focused laser for photon generation and qubit control},
	url = {https://quantumoptics.at/images/publications/diploma/master_Marco_Canteri.pdf},
	year = {2020}}

@phdthesis{DarioThesis,
	author = {Dario Alessandro Fioretto},
	langid = {english},
	school = {University of Innsbruck},
	title = {Towards a flexible source for indistinguishable photons based on trapped ions and cavities},
	url = {https://diglib.uibk.ac.at/ulbtirolhs/download/pdf/5459504}}

@article{CHOI1975285,
	author = {Man-Duen Choi},
	doi = {https://doi.org/10.1016/0024-3795(75)90075-0},
	issn = {0024-3795},
	journal = {Linear Algebra and its Applications},
	number = {3},
	pages = {285-290},
	title = {Completely positive linear maps on complex matrices},
	url = {https://www.sciencedirect.com/science/article/pii/0024379575900750},
	volume = {10},
	year = {1975}}

@article{JAMIOLKOWSKI1972275,
	author = {A. Jamio{\l}kowski},
	doi = {https://doi.org/10.1016/0034-4877(72)90011-0},
	issn = {0034-4877},
	journal = {Reports on Mathematical Physics},
	number = {4},
	pages = {275-278},
	title = {Linear transformations which preserve trace and positive semidefiniteness of operators},
	url = {https://www.sciencedirect.com/science/article/pii/0034487772900110},
	volume = {3},
	year = {1972}}

@mastersthesis{JohannesThesis,
	author = {Johannes Helgert},
	langid = {English},
	school = {University of Innsbruck},
	title = {Radio frequency control of a trapped-ion quantum network node},
	year = {2024}}

@article{PhysRevA.71.062310,
	author = {Gilchrist, Alexei and Langford, Nathan K. and Nielsen, Michael A.},
	doi = {10.1103/PhysRevA.71.062310},
	issue = {6},
	journal = {Phys. Rev. A},
	month = {Jun},
	numpages = {14},
	pages = {062310},
	publisher = {American Physical Society},
	title = {Distance measures to compare real and ideal quantum processes},
	url = {https://link.aps.org/doi/10.1103/PhysRevA.71.062310},
	volume = {71},
	year = {2005}}

@article{NetworkPaper,
	author = {Krutyanskiy, V. and Galli, M. and Krcmarsky, V. and Baier, S. and Fioretto, D. A. and Pu, Y. and Mazloom, A. and Sekatski, P. and Canteri, M. and Teller, M. and Schupp, J. and Bate, J. and Meraner, M. and Sangouard, N. and Lanyon, B. P. and Northup, T. E.},
	doi = {10.1103/PhysRevLett.130.050803},
	issue = {5},
	journal = {Phys. Rev. Lett.},
	month = {Feb},
	numpages = {7},
	pages = {050803},
	publisher = {American Physical Society},
	title = {Entanglement of Trapped-Ion Qubits Separated by 230 Meters},
	url = {https://link.aps.org/doi/10.1103/PhysRevLett.130.050803},
	volume = {130},
	year = {2023}}

@article{JosefPaper,
	author = {Schupp, J. and Krcmarsky, V. and Krutyanskiy, V. and Meraner, M. and Northup, T.E. and Lanyon, B.P.},
	doi = {10.1103/PRXQuantum.2.020331},
	issue = {2},
	journal = {PRX Quantum},
	month = {Jun},
	numpages = {16},
	pages = {020331},
	publisher = {American Physical Society},
	title = {Interface between Trapped-Ion Qubits and Traveling Photons with Close-to-Optimal Efficiency},
	url = {https://link.aps.org/doi/10.1103/PRXQuantum.2.020331},
	volume = {2},
	year = {2021}}

@article{Stute2013,
	author = {Stute, A. and Casabone, B. and Brandst{\"a}tter, B. and Friebe, K. and Northup, T. E. and Blatt, R.},
	day = {01},
	doi = {10.1038/nphoton.2012.358},
	issn = {1749-4893},
	journal = {Nature Photonics},
	month = {Mar},
	number = {3},
	pages = {219-222},
	title = {Quantum-state transfer from an ion to a photon},
	url = {https://doi.org/10.1038/nphoton.2012.358},
	volume = {7},
	year = {2013}}

@article{PhysRevLett.130.090803,
	author = {Drmota, P. and Main, D. and Nadlinger, D. P. and Nichol, B. C. and Weber, M. A. and Ainley, E. M. and Agrawal, A. and Srinivas, R. and Araneda, G. and Ballance, C. J. and Lucas, D. M.},
	doi = {10.1103/PhysRevLett.130.090803},
	issue = {9},
	journal = {Phys. Rev. Lett.},
	month = {Mar},
	numpages = {7},
	pages = {090803},
	publisher = {American Physical Society},
	title = {Robust Quantum Memory in a Trapped-Ion Quantum Network Node},
	url = {https://link.aps.org/doi/10.1103/PhysRevLett.130.090803},
	volume = {130},
	year = {2023}}

@article{RevModPhys.82.1209,
	author = {Duan, L.-M. and Monroe, C.},
	doi = {10.1103/RevModPhys.82.1209},
	issue = {2},
	journal = {Rev. Mod. Phys.},
	month = {Apr},
	numpages = {0},
	pages = {1209--1224},
	publisher = {American Physical Society},
	title = {Colloquium: Quantum networks with trapped ions},
	url = {https://link.aps.org/doi/10.1103/RevModPhys.82.1209},
	volume = {82},
	year = {2010}}

@article{PhysRevLett.124.013602,
	author = {Takahashi, Hiroki and Kassa, Ezra and Christoforou, Costas and Keller, Matthias},
	doi = {10.1103/PhysRevLett.124.013602},
	issue = {1},
	journal = {Phys. Rev. Lett.},
	month = {Jan},
	numpages = {5},
	pages = {013602},
	publisher = {American Physical Society},
	title = {Strong Coupling of a Single Ion to an Optical Cavity},
	url = {https://link.aps.org/doi/10.1103/PhysRevLett.124.013602},
	volume = {124},
	year = {2020}}

@article{Stute2012,
	author = {Stute, A. and Casabone, B. and Schindler, P. and Monz, T. and Schmidt, P. O. and Brandst{\"a}tter, B. and Northup, T. E. and Blatt, R.},
	day = {01},
	doi = {10.1038/nature11120},
	issn = {1476-4687},
	journal = {Nature},
	month = {May},
	number = {7399},
	pages = {482-485},
	title = {Tunable ion--photon entanglement in an optical cavity},
	url = {https://doi.org/10.1038/nature11120},
	volume = {485},
	year = {2012}}

@article{Hucul2015,
	author = {Hucul, D. and Inlek, I. V. and Vittorini, G. and Crocker, C. and Debnath, S. and Clark, S. M. and Monroe, C.},
	day = {01},
	doi = {10.1038/nphys3150},
	issn = {1745-2481},
	journal = {Nature Physics},
	month = {Jan},
	number = {1},
	pages = {37-42},
	title = {Modular entanglement of atomic qubits using photons and phonons},
	url = {https://doi.org/10.1038/nphys3150},
	volume = {11},
	year = {2015}}

@article{PhysRevLett.67.661,
	author = {Ekert, Artur K.},
	doi = {10.1103/PhysRevLett.67.661},
	issue = {6},
	journal = {Phys. Rev. Lett.},
	month = {Aug},
	numpages = {0},
	pages = {661--663},
	publisher = {American Physical Society},
	title = {Quantum cryptography based on Bell's theorem},
	url = {https://link.aps.org/doi/10.1103/PhysRevLett.67.661},
	volume = {67},
	year = {1991}}

@article{ultrahighfidelity,
	author = {Gevorgyan, Hayk L. and Vitanov, Nikolay V.},
	doi = {10.1103/PhysRevA.104.012609},
	issue = {1},
	journal = {Phys. Rev. A},
	month = {Jul},
	numpages = {12},
	pages = {012609},
	publisher = {American Physical Society},
	title = {Ultrahigh-fidelity composite rotational quantum gates},
	url = {https://link.aps.org/doi/10.1103/PhysRevA.104.012609},
	volume = {104},
	year = {2021}}

@article{PhysRevLett.132.150604,
	author = {Drmota, P. and Nadlinger, D. P. and Main, D. and Nichol, B. C. and Ainley, E. M. and Leichtle, D. and Mantri, A. and Kashefi, E. and Srinivas, R. and Araneda, G. and Ballance, C. J. and Lucas, D. M.},
	doi = {10.1103/PhysRevLett.132.150604},
	issue = {15},
	journal = {Phys. Rev. Lett.},
	month = {Apr},
	numpages = {6},
	pages = {150604},
	publisher = {American Physical Society},
	title = {Verifiable Blind Quantum Computing with Trapped Ions and Single Photons},
	url = {https://link.aps.org/doi/10.1103/PhysRevLett.132.150604},
	volume = {132},
	year = {2024}}

@article{Giovannetti2011,
	author = {Giovannetti, Vittorio and Lloyd, Seth and Maccone, Lorenzo},
	day = {01},
	doi = {10.1038/nphoton.2011.35},
	issn = {1749-4893},
	journal = {Nature Photonics},
	month = {Apr},
	number = {4},
	pages = {222-229},
	title = {Advances in quantum metrology},
	url = {https://doi.org/10.1038/nphoton.2011.35},
	volume = {5},
	year = {2011}}

@article{Marciniak:2021tld,
	archiveprefix = {arXiv},
	author = {Marciniak, Christian D. and Feldker, Thomas and Pogorelov, Ivan and Kaubruegger, Raphael and Vasilyev, Denis V. and van Bijnen, Rick and Schindler, Philipp and Zoller, Peter and Blatt, Rainer and Monz, Thomas},
	doi = {10.1038/s41586-022-04435-4},
	eprint = {2107.01860},
	journal = {Nature},
	number = {7902},
	pages = {604--609},
	primaryclass = {quant-ph},
	title = {{Optimal metrology with programmable quantum sensors}},
	volume = {603},
	year = {2022}}

@article{James1998,
	author = {James, D. F. V.},
	day = {01},
	doi = {10.1007/s003400050373},
	issn = {1432-0649},
	journal = {Applied Physics B},
	month = {Feb},
	number = {2},
	pages = {181-190},
	title = {Quantum dynamics of cold trapped ions with application to quantum computation},
	url = {https://doi.org/10.1007/s003400050373},
	volume = {66},
	year = {1998}}

@article{deLange:2010avk,
	archiveprefix = {arXiv},
	author = {de Lange, G. and Wang, Z. H. and Rist{\`e}, D. and Dobrovitski, V. V. and Hanson, R.},
	doi = {10.1126/science.1192739},
	eprint = {1008.2119},
	journal = {Science},
	number = {6000},
	pages = {1192739},
	primaryclass = {quant-ph},
	title = {{Universal Dynamical Decoupling of a Single Solid-State Spin from a Spin Bath}},
	volume = {330},
	year = {2010}}

@article{doi:10.1098/rsta.2011.0355,
	author = {Souza, Alexandre M. and {\'A}lvarez, Gonzalo A. and Suter, Dieter},
	doi = {10.1098/rsta.2011.0355},
	journal = {Philosophical Transactions of the Royal Society A: Mathematical, Physical and Engineering Sciences},
	number = {1976},
	pages = {4748-4769},
	title = {Robust dynamical decoupling},
	volume = {370},
	year = {2012}}

@article{Fitzsimons:2016xat,
	archiveprefix = {arXiv},
	author = {Fitzsimons, Joseph F.},
	doi = {10.1038/s41534-017-0025-3},
	eprint = {1611.10107},
	journal = {npj Quantum Inf.},
	number = {1},
	pages = {23},
	primaryclass = {quant-ph},
	title = {{Private quantum computation: an introduction to blind quantum computing and related protocols}},
	volume = {3},
	year = {2017}}

@article{dur,
	adsurl = {https://ui.adsabs.harvard.edu/abs/2020PhRvR...2b3052S},
	archiveprefix = {arXiv},
	author = {{Sekatski}, P. and {W{\"o}lk}, S. and {D{\"u}r}, W.},
	doi = {10.1103/PhysRevResearch.2.023052},
	eid = {023052},
	eprint = {1905.06765},
	journal = {Physical Review Research},
	month = apr,
	number = {2},
	pages = {023052},
	primaryclass = {quant-ph},
	title = {{Optimal distributed sensing in noisy environments}},
	volume = {2},
	year = 2020}

@article{networkclocks,
	adsurl = {https://ui.adsabs.harvard.edu/abs/2014NatPh..10..582K},
	archiveprefix = {arXiv},
	author = {{K{\'o}m{\'a}r}, P. and {Kessler}, E.~M. and {Bishof}, M. and {Jiang}, L. and {S{\o}rensen}, A.~S. and {Ye}, J. and {Lukin}, M.~D.},
	doi = {10.1038/nphys3000},
	eprint = {1310.6045},
	journal = {Nature Physics},
	month = aug,
	number = {8},
	pages = {582-587},
	primaryclass = {quant-ph},
	title = {{A quantum network of clocks}},
	volume = {10},
	year = 2014}

@article{PhysRevA.89.022317,
	author = {Monroe, C. and Raussendorf, R. and Ruthven, A. and Brown, K. R. and Maunz, P. and Duan, L.-M. and Kim, J.},
	doi = {10.1103/PhysRevA.89.022317},
	issue = {2},
	journal = {Phys. Rev. A},
	month = {Feb},
	numpages = {16},
	pages = {022317},
	publisher = {American Physical Society},
	title = {Large-scale modular quantum-computer architecture with atomic memory and photonic interconnects},
	url = {https://link.aps.org/doi/10.1103/PhysRevA.89.022317},
	volume = {89},
	year = {2014}}

@article{Knaut2024,
	author = {Knaut, C. M. and Suleymanzade, A. and Wei, Y.-C. and Assumpcao, D. R. and Stas, P.-J. and Huan, Y. Q. and Machielse, B. and Knall, E. N. and Sutula, M. and Baranes, G. and Sinclair, N. and De-Eknamkul, C. and Levonian, D. S. and Bhaskar, M. K. and Park, H. and Lon{\v{c}}ar, M. and Lukin, M. D.},
	day = {01},
	doi = {10.1038/s41586-024-07252-z},
	issn = {1476-4687},
	journal = {Nature},
	month = {May},
	number = {8012},
	pages = {573-578},
	title = {Entanglement of nanophotonic quantum memory nodes in a telecom network},
	url = {https://doi.org/10.1038/s41586-024-07252-z},
	volume = {629},
	year = {2024}}

@article{PhysRevLett.70.3213,
	author = {Jin, Jian and Church, D. A.},
	doi = {10.1103/PhysRevLett.70.3213},
	issue = {21},
	journal = {Phys. Rev. Lett.},
	month = {May},
	numpages = {0},
	pages = {3213--3216},
	publisher = {American Physical Society},
	title = {Precision lifetimes for the Ca+4p2P levels: Experiment challenges theory at the 1 percent level},
	url = {https://link.aps.org/doi/10.1103/PhysRevLett.70.3213},
	volume = {70},
	year = {1993}}

@article{Cui:2025olz,
	archiveprefix = {arXiv},
	author = {Cui, Z. B. and others},
	eprint = {2510.20392},
	journal = {arXiv},
	month = {10},
	primaryclass = {quant-ph},
	title = {{Multiplexed ion-ion entanglement over $1.2$ kilometer fibers}},
	year = {2025}}

@article{Stolk:2024xop,
	archiveprefix = {arXiv},
	author = {Stolk, Arian J. and others},
	doi = {10.1126/sciadv.adp6442},
	eprint = {2404.03723},
	journal = {Sci. Adv.},
	number = {44},
	pages = {adp6442},
	primaryclass = {quant-ph},
	title = {{Metropolitan-scale heralded entanglement of solid-state qubits}},
	volume = {10},
	year = {2024}}

@article{vanLeent2022,
	author = {van Leent, Tim and Bock, Matthias and Fertig, Florian and Garthoff, Robert and Eppelt, Sebastian and Zhou, Yiru and Malik, Pooja and Seubert, Matthias and Bauer, Tobias and Rosenfeld, Wenjamin and Zhang, Wei and Becher, Christoph and Weinfurter, Harald},
	day = {01},
	doi = {10.1038/s41586-022-04764-4},
	issn = {1476-4687},
	journal = {Nature},
	month = {Jul},
	number = {7917},
	pages = {69-73},
	title = {Entangling single atoms over 33{\thinspace}km telecom fibre},
	url = {https://doi.org/10.1038/s41586-022-04764-4},
	volume = {607},
	year = {2022}}

@article{Saha2025,
	author = {Saha, Sagnik and Shalaev, Mikhail and O'Reilly, Jameson and Goetting, Isabella and Toh, George and Kalakuntla, Ashish and Yu, Yichao and Monroe, Christopher},
	day = {14},
	doi = {10.1038/s41467-025-57557-4},
	issn = {2041-1723},
	journal = {Nature Communications},
	month = {Mar},
	number = {1},
	pages = {2533},
	title = {High-fidelity remote entanglement of trapped atoms mediated by time-bin photons},
	url = {https://doi.org/10.1038/s41467-025-57557-4},
	volume = {16},
	year = {2025}}

@article{PhysRevLett.119.010503,
	author = {Stockill, R. and Stanley, M. J. and Huthmacher, L. and Clarke, E. and Hugues, M. and Miller, A. J. and Matthiesen, C. and Le Gall, C. and Atat\"ure, M.},
	doi = {10.1103/PhysRevLett.119.010503},
	issue = {1},
	journal = {Phys. Rev. Lett.},
	month = {Jul},
	numpages = {6},
	pages = {010503},
	publisher = {American Physical Society},
	title = {Phase-Tuned Entangled State Generation between Distant Spin Qubits},
	url = {https://link.aps.org/doi/10.1103/PhysRevLett.119.010503},
	volume = {119},
	year = {2017}}

@article{Ritter2012,
	author = {Ritter, Stephan and N{\"o}lleke, Christian and Hahn, Carolin and Reiserer, Andreas and Neuzner, Andreas and Uphoff, Manuel and M{\"u}cke, Martin and Figueroa, Eden and Bochmann, Joerg and Rempe, Gerhard},
	day = {01},
	doi = {10.1038/nature11023},
	issn = {1476-4687},
	journal = {Nature},
	month = {Apr},
	number = {7393},
	pages = {195-200},
	title = {An elementary quantum network of single atoms in optical cavities},
	url = {https://doi.org/10.1038/nature11023},
	volume = {484},
	year = {2012}}

@article{Kimble2008,
	author = {Kimble, H. J.},
	day = {01},
	doi = {10.1038/nature07127},
	issn = {1476-4687},
	journal = {Nature},
	month = {Jun},
	number = {7198},
	pages = {1023-1030},
	title = {The quantum internet},
	url = {https://doi.org/10.1038/nature07127},
	volume = {453},
	year = {2008}}

@article{doi:10.1126/science.aam9288,
	author = {Stephanie Wehner and David Elkouss and Ronald Hanson},
	doi = {10.1126/science.aam9288},
	journal = {Science},
	number = {6412},
	pages = {eaam9288},
	title = {Quantum internet: A vision for the road ahead},
	volume = {362},
	year = {2018}}

@article{shuttling,
	author = {Canteri, M. and Koong, Z. X. and Bate, J. and Winkler, A. and Krutyanskiy, V. and Lanyon, B. P.},
	doi = {10.1103/v5k1-whwz},
	issue = {8},
	journal = {Phys. Rev. Lett.},
	month = {Aug},
	numpages = {7},
	pages = {080801},
	publisher = {American Physical Society},
	title = {Photon-Interfaced Ten-Qubit Register of Trapped Ions},
	url = {https://link.aps.org/doi/10.1103/v5k1-whwz},
	volume = {135},
	year = {2025}}

@article{PhysRevLett.125.260502,
	author = {Magnard, P. and Storz, S. and Kurpiers, P. and Sch\"ar, J. and Marxer, F. and L\"utolf, J. and Walter, T. and Besse, J.-C. and Gabureac, M. and Reuer, K. and Akin, A. and Royer, B. and Blais, A. and Wallraff, A.},
	doi = {10.1103/PhysRevLett.125.260502},
	issue = {26},
	journal = {Phys. Rev. Lett.},
	month = {Dec},
	numpages = {7},
	pages = {260502},
	publisher = {American Physical Society},
	title = {Microwave Quantum Link between Superconducting Circuits Housed in Spatially Separated Cryogenic Systems},
	url = {https://link.aps.org/doi/10.1103/PhysRevLett.125.260502},
	volume = {125},
	year = {2020}}

@article{Hensen2015,
	author = {Hensen, B. and Bernien, H. and Dr{\'e}au, A. E. and Reiserer, A. and Kalb, N. and Blok, M. S. and Ruitenberg, J. and Vermeulen, R. F. L. and Schouten, R. N. and Abell{\'a}n, C. and Amaya, W. and Pruneri, V. and Mitchell, M. W. and Markham, M. and Twitchen, D. J. and Elkouss, D. and Wehner, S. and Taminiau, T. H. and Hanson, R.},
	day = {01},
	doi = {10.1038/nature15759},
	issn = {1476-4687},
	journal = {Nature},
	month = {Oct},
	number = {7575},
	pages = {682-686},
	title = {Loophole-free Bell inequality violation using electron spins separated by 1.3 kilometres},
	url = {https://doi.org/10.1038/nature15759},
	volume = {526},
	year = {2015}}

@article{doi:10.1126/science.abg1919,
	author = {M. Pompili and S. L. N. Hermans and S. Baier and H. K. C. Beukers and P. C. Humphreys and R. N. Schouten and R. F. L. Vermeulen and M. J. Tiggelman and L. dos Santos Martins and B. Dirkse and S. Wehner and R. Hanson},
	doi = {10.1126/science.abg1919},
	journal = {Science},
	number = {6539},
	pages = {259-264},
	title = {Realization of a multinode quantum network of remote solid-state qubits},
	volume = {372},
	year = {2021}}

\end{document}